\documentclass[11pt]{article}

\usepackage[final]{acl}

\usepackage{times}
\usepackage{latexsym}
\usepackage[T1]{fontenc}
\usepackage[utf8]{inputenc}
\usepackage{microtype}
\usepackage{inconsolata}
\usepackage{graphicx}
\usepackage{enumitem}
\usepackage[utf8]{inputenc} 
\usepackage[T1]{fontenc}    
\usepackage{wrapfig}  

\usepackage{hyperref}       
\usepackage{url}            
\usepackage{booktabs}       
\usepackage{amsfonts}       
\usepackage{nicefrac}       
\usepackage{graphicx}
\usepackage{microtype}      
\usepackage{xcolor}         
\usepackage{amsmath} 
\usepackage{multirow} 
\usepackage{cleveref}
\usepackage{listings}
\usepackage{listingsutf8}

\newcommand{\horizon}{\texttt{HORIZON}}

\title{\texttt{HORIZON}: A Benchmark for in-the-wild User Behavior Modelling}
\usepackage{tcolorbox}
\tcbuselibrary{listingsutf8}

\newtcblisting{promptbox}{
  colback=gray!5,
  colframe=gray!80,
  listing only,
  listing options={
    basicstyle=\ttfamily\scriptsize,
    breaklines=true,
    columns=fullflexible,
    showstringspaces=false,
  },
  left=2mm,
  right=2mm,
  top=1mm,
  bottom=1mm,
  boxrule=0.5pt,
  arc=2pt
}

\author{
  \parbox{0.9\linewidth}{\centering 
    Arnav Goel$^{\spadesuit}$\thanks{Work done while at Microsoft Research India} \quad 
    Pranjal A. Chitale$^{\heartsuit}$ \quad 
    Bhawna Paliwal$^{\diamondsuit}$\footnotemark[1] \\[0.3em]
    Bishal Santra$^{\heartsuit}$ \quad
    Amit Sharma$^{\heartsuit}$\thanks{Corresponding author} \\[0.5em]
    $^{\spadesuit}${\rm Carnegie Mellon University} \quad 
    $^{\heartsuit}${\rm Microsoft Research India} \quad \\
    $^{\diamondsuit}${\rm University of California, Berkeley} \\
    {\tt \small arnavgoe@cs.cmu.edu, amshar@microsoft.com} \\ [0.2em]
    \small
    \raisebox{-0.5ex}{\includegraphics[height=0.8em]{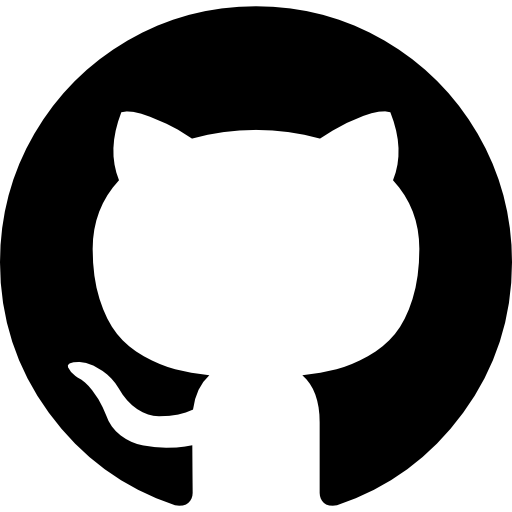}}~
    \href{https://aka.ms/horizon-benchmark}{microsoft/horizon-benchmark}
    \quad
    }
}

\begin{document}
\maketitle
\begin{abstract} 
User behavior in the real world is diverse, cross-domain, and spans long time horizons. Existing user modeling benchmarks however remain narrow, focusing mainly on short sessions and next-item prediction within a single domain. Such limitations hinder progress toward robust and generalizable user models. We present \horizon\, a new benchmark that reformulates user modeling along three axes i.e.\ dataset, task, and evaluation. Built from a large-scale, cross-domain reformulation of Amazon Reviews, \horizon\ covers 54M users and 35M items, enabling both pretraining and realistic evaluation of models in heterogeneous environments. Unlike prior benchmarks, it challenges models to generalize across domains, users, and time, moving beyond standard missing-positive prediction in the same domain. We propose new tasks and evaluation setups that better reflect real-world deployment scenarios. These include temporal generalization, sequence-length variation, and modeling unseen users, with metrics designed to assess general user behavior understanding rather than isolated next-item prediction. We benchmark popular sequential recommendation architectures alongside LLM-based baselines that leverage long-term interaction histories. Our results highlight the gap between current methods and the demands of real-world user modeling, while establishing \horizon\ as a foundation for research on temporally robust, cross-domain, and general-purpose user models.
\end{abstract}

\section{Introduction}
\label{sec:introduction}
Personalization is at the core of modern digital platforms, driving user engagement in domains such as e-commerce, streaming, and social networks by tailoring content and services to individual preferences. Early personalization methods relied mainly on static user representations and focused on prediction of next-item within single-domain datasets such as MovieLens~\citep{harper2015movielens} and Amazon Reviews~\citep{ni2019justifying}. However, contemporary user behavior is inherently multi-faceted, spanning diverse content types and multiple platforms, and reflecting complex, evolving preferences and latent interests that are not adequately captured by static models \citep{zhou2024use,treves2025viki,hou2022towards,lin2019cross}. To address these challenges, recent work has increasingly framed personalization as a \textit{sequential recommendation} problem, aiming to model long-term dependencies and dynamic user intent from interaction histories~\citep{SASRec,BERT4Rec,hou2022towards}.

Despite significant advances in modeling world knowledge~\citep{yue2023llamarec,wang2024can,goel2023advancementsscientificcontrollabletext} and semantic reasoning driven by transformers and pretrained large language models (LLMs)~\citep{10.1007/978-3-031-49601-1_4, 10.1145/3728483.3760194, li2025attributingcultureconditionedgenerationspretraining}, prior work in personalization largely remains constrained to \textit{single-domain} recommendation settings. Such formulations fail to leverage cross-domain signals and instead learn fragmented, domain-specific representations of users, capturing only a partial view of their preferences \citep{SASRec,BERT4Rec,hou2022towards,hou2024bridging}.
Even when recent datasets, such as Amazon-Reviews 2023 \citep{hou2024bridging}, incorporate interactions across multiple domains, evaluation protocols and task formulations continue to treat each domain independently. This disconnect highlights a critical gap between real-world user behavior and existing benchmarking practices, motivating the need for a unified framework that can evaluate and advance truly cross-domain personalization.

\begin{table*}[t]
\centering
\label{tab:all_bench_comparison}
\small
\begin{tabular}{lccccc}
\toprule
\textbf{Attribute} & \textbf{PF} & \textbf{Amz-M2} & \textbf{MIND} & \textbf{Amz-Reviews} & \textbf{\texttt{HORIZON}} \\
\midrule
No. of users & N/A & N/A & 1M & 54.51M & \textbf{54.51M} \\
Avg User History Length & N/A & 4.2 & N/A & 3.86 & \textbf{9.07} \\
No. of items & N/A & 1.42M & 0.16M & 34.52M & \textbf{34.52M} \\
No. of interactions & N/A & 16.78M & 24.15M& 485.89M & \textbf{485.89M} \\
\midrule
Cross-domain & \checkmark & \checkmark & $\times$ & $\times$ & \checkmark \\
Diversity & \checkmark & $\times$ & $\times$ & $\times$ & \checkmark \\
Interaction Timestamps & -- & $\times$ & $\times$ & \checkmark & \checkmark \\
Open-Source & $\times$ & \checkmark & \checkmark & \checkmark & \checkmark \\
\bottomrule
\end{tabular}
\caption{Comparison of existing Sequential Recommendation Benchmarks with \texttt{HORIZON}. (\texttt{PF} refers to PinnerFormer, \texttt{Amz-M2} refers to Amazon M2, \texttt{Amz-Reviews} is the Amazon Reviews dataset.)}
\end{table*}


Existing benchmarks predominantly focus on single-domain, next-item prediction, inadvertently encouraging models to exploit item--item similarities rather than develop a holistic understanding of user preferences~\citep{rendle2020neural}. While recent efforts such as PinnerFormer \citep{PinnerFormer} and USE \citep{zhou2024use} highlight the importance of richer user modeling, they rely on private datasets, limiting reproducibility and leaving a critical gap in open benchmarking standards.
Moreover, real-world recommendation scenarios often unfold over extended time horizons. For instance, user behaviors in e-commerce require models to reason over multi-year interaction histories and anticipate long-term intent. Such capabilities are essential for applications like proactive recommendation and inventory planning. However, current benchmarks, constrained to short-term next-item prediction within limited temporal windows, fail to evaluate whether models can capture long-range dependencies and evolving user preferences across extended time spans.

Taken together, these limitations point to a fundamental gap in current evaluation practices: existing benchmarks do not adequately measure whether models can generalize across domains, reason over long temporal horizons, or capture deeper semantic structure in user behavior. In particular, three key challenges remain underexplored: \textbf{(i) cross-domain generalization}, where models must transfer knowledge across diverse content domains and platforms; \textbf{(ii) long-range temporal generalization}, requiring anticipation of user intent far beyond the immediate training window; and \textbf{(iii) semantic understanding}, involving the ability to uncover latent, non-obvious relationships within user interaction histories.

To address these challenges, we introduce \texttt{HORIZON}, a fully open-source, large-scale benchmark for evaluating sequential recommendation models under realistic cross-domain and long-horizon settings. Unlike prior benchmarks that rely on ratio-based temporal splits and short-term evaluation, \texttt{HORIZON} disentangles key generalization axes by explicitly distinguishing between \textit{seen} and \textit{unseen} users, as well as temporally \textit{close} versus \textit{distant} scenarios relative to the training distribution. This structured evaluation reveals insights obscured by existing paradigms: we observe substantial variation in model performance across temporal and user generalization regimes, showing that strong in-distribution results do not reliably translate to real-world robustness. For instance, BERT4Rec~\citep{BERT4Rec}, while state-of-the-art in standard settings, degrades significantly for out-of-distribution users yet remains competitive in long-range temporal extrapolation (e.g., Recall@50 and Recall@100). Moreover, we find that LLMs do not consistently outperform specialized architectures on user behavior modeling tasks.

\noindent\textbf{In summary, our contributions are as follows:}
\begin{enumerate}[noitemsep]
    \item We introduce \texttt{HORIZON}, the first fully open-source benchmark for evaluating sequential recommendation models across \textit{cross-domain} and \textit{long-horizon} personalization settings.
    \item We propose a unified evaluation framework that disentangles key generalization axes i.e. \textit{seen vs.\ unseen users} and \textit{seen vs.\ unseen timeframes}, enabling more faithful and fine-grained assessment of real-world performance.
    \item We provide a comprehensive empirical study revealing previously unobserved trade-offs in modern recommendation models, including discrepancies between in-distribution accuracy, temporal generalization, and out-of-distribution user adaptation, as well as the limitations of LLMs for user modeling.
\end{enumerate}

\section{Related Work}
\label{sec:related_work}

\paragraph{Sequential Recommendation Datasets:}
Sequential recommendation research has primarily relied on a small number of established benchmarks that, while influential, offer limited coverage of realistic user behavior. MovieLens~\citep{movielens} remains one of the most widely used datasets, providing temporally ordered movie ratings; however, it is relatively small in scale and strictly single-domain. Even its largest variant, MovieLens-25M, contains only 25 million interactions and reflects a narrow media-focused item space, limiting its suitability for evaluating large-scale or cross-domain user modeling.

Several other commonly used datasets share similar constraints. Yelp, which contains approximately 1.2 million users and 5 million reviews of local businesses,\footnote{\url{https://business.yelp.com/data/resources/open-dataset/}} and Gowalla~\citep{10.1145/2020408.2020579}, a location-based dataset with 197K users and 6.4M check-ins collected between 2009-2012, are both single-domain and exhibit weak sequential structure. Recent analysis shows that shuffling test-time sequences in such datasets has minimal impact on the  performance, suggesting limited temporal dependency \citep{klenitskiy2024we}. Similarly, the Steam dataset \citep{steamdataset}, while capturing richer behavioral signals such as purchases and playtime, remains confined to the gaming domain and is modest in scale, with fewer than 8 million interactions.

The Amazon Reviews dataset~\citep{ni2019justifying, hou2024bridging} provides substantially larger coverage across many product categories, but is typically evaluated by partitioning categories into isolated recommendation tasks. This segmentation prevents models from capturing cross-category transitions that naturally arise in real user behavior and often leads to severe sparsity within individual categories. As a result, existing benchmarks fall short of supporting large-scale, cross-domain, and temporally grounded evaluation of sequential user modeling.

\paragraph{Towards Large-scale and Long Horizon User Modeling:}
To address the limitations of traditional benchmarks, several datasets have sought to capture richer and broader user behaviors. MIND \citep{wu-etal-2020-mind} provides large-scale news consumption logs with approximately one million users and rich textual features, but is restricted to a single domain and short user histories spanning only two weeks, limiting its suitability for long-horizon or cross-domain evaluation. Amazon-M2 \citep{amazon-m2} extends coverage to multilingual and cross-locale e-commerce interactions across six regions, but is primarily designed for session-based recommendation rather than modeling long-term user behavior over extended time spans.

More recent efforts approach the required scale and temporal depth but remain inaccessible. The Pinterest dataset used in PinnerFormer \citep{PinnerFormer} contains billions of multimodal interactions across multiple years, and USE \citep{zhou2024use} includes diverse behavioral sequences from Snapchat users. However, both datasets are proprietary, limiting reproducibility and preventing their adoption as public benchmarks for large-scale, long-horizon user modeling.

\section{\horizon\ Benchmark}
\label{sec:dataset_comparison}
\noindent\textbf{Benchmark Description}:
User modeling and sequential recommendation aim to predict a user's future interactions based on their past behavior. Formally, for a user $u$, we observe a sequence of interactions over time $\mathcal{H}_u = [i_1, i_2, \dots, i_t]$, where $i_t$ denotes the item interacted with at time $t$. The objective is to estimate the likelihood of the next interaction $i_{t+1}$ or future next interactions over some time period $T$ $i.e.$ $\hat{i}_{t+1,..,T} = (i_{t+1}, i_{t+2}, .... i_{T})$, given the user's historical context:
\[
\hat{i}_{t+1} = \arg\max_{i \in \mathcal{I}} \Pr(i \mid \mathcal{H}_u),
\]
where $\mathcal{I}$ denotes the candidate item set. This formulation underpins several established benchmarks such as MIND, M2, and Amazon Reviews \citep{wu2020mind, jin2023amazon, hou2024bridging}.
As noted in \Cref{sec:related_work}, the Amazon Reviews dataset has become a widely used resource for training and evaluating sequential recommenders. However, it segregates user interactions by product categories, making it domain-specific and thus limiting its ability to capture holistic user preferences. In the real world, users engage with a variety of domains, and isolating interactions to a single domain introduces artificial boundaries, resulting in incomplete modeling of cross-domain behaviors and potentially spurious patterns causing incorrect user modeling.

To address this limitation, we introduce \texttt{HORIZON}, a large-scale benchmark designed to support cross-domain user modeling and sequential recommendation. \horizon\ is constructed by refactoring and consolidating the Amazon Reviews 2023 dataset \citep{hou2024bridging}, merging interactions across all available categories to create unified, realistic user histories. The resulting benchmark comprises of 53.5 million users and 34.5 million unique items, enabling rigorous evaluation of models under settings that better reflect real-world recommendation scenarios. \footnote{Detailed plots and stats for the benchmark are added to the appendix \ref{appendix:A}}

\subsection{Comparison with Existing Benchmarks}
Table \Cref{tab:all_bench_comparison} provides a comparative analysis of our dataset against existing sequential recommendation benchmarks. While proprietary datasets like PinnerFormer ~\citep{PinnerFormer} offer scale and diversity, they remain inaccessible to the broader research community. Public datasets such as Amazon-M2 ~\citep{amazon-m2} provide cross-domain capabilities but lack temporal depth due to these being restricted to session-based interactions rather than long-term user modeling. The MIND dataset~\citep{wu-etal-2020-mind}, despite its million-user scale, covers only two weeks of user history, severely limiting its utility for long-horizon recommendation research ~\citep{klenitskiy2024we}. Similarly, the Amazon Reviews dataset~\citep{ni2019justifying, hou2024bridging} provides timestamps, but artificially segments interactions into isolated domains. In contrast, \texttt{HORIZON} uniquely combines cross-domain coverage, interaction diversity, and comprehensive temporal information, enabling more realistic evaluation of sequential recommendation systems across extended time horizons.



\section{Task Formulations}
\label{sec:eval_methods}
\subsection{Traditional Evaluation Setups}
\paragraph{Drawbacks of Traditional Evaluation Setups}

Traditional evaluation methodologies in recommendation systems have primarily focused on \textbf{in-distribution settings}, where the training, validation, and test splits are derived from the same user interaction trace. This leads to substantial overlap in distributional characteristics across splits, limiting the capacity to evaluate generalization or robustness in real-world scenarios \citep{mcelfresh2022generalizability}. Two widely adopted paradigms for this evaluation are \textit{Leave-One-Out} and \textit{Ratio-Based} evaluations.

\noindent\textbf{Leave-One-Out.} For a user history sequence with $n$ events, the $(n{-}1)^\text{th}$ interaction is held out for validation, the $n^\text{th}$ for testing, and the preceding $(n{-}2)$ interactions form the training set. This approach has been widely deployed for evaluating user modeling architectures in recent years \citep{sun2023take} but can often leak future interactions into training data, violating the temporal order of real-world scenarios. This leads to inflated performance metrics that don’t reflect practical deployment conditions \citep{10.1145/3383313.3418479, ji2023critical}.

\begin{figure}
  \centering
  \includegraphics[width=0.48\textwidth]{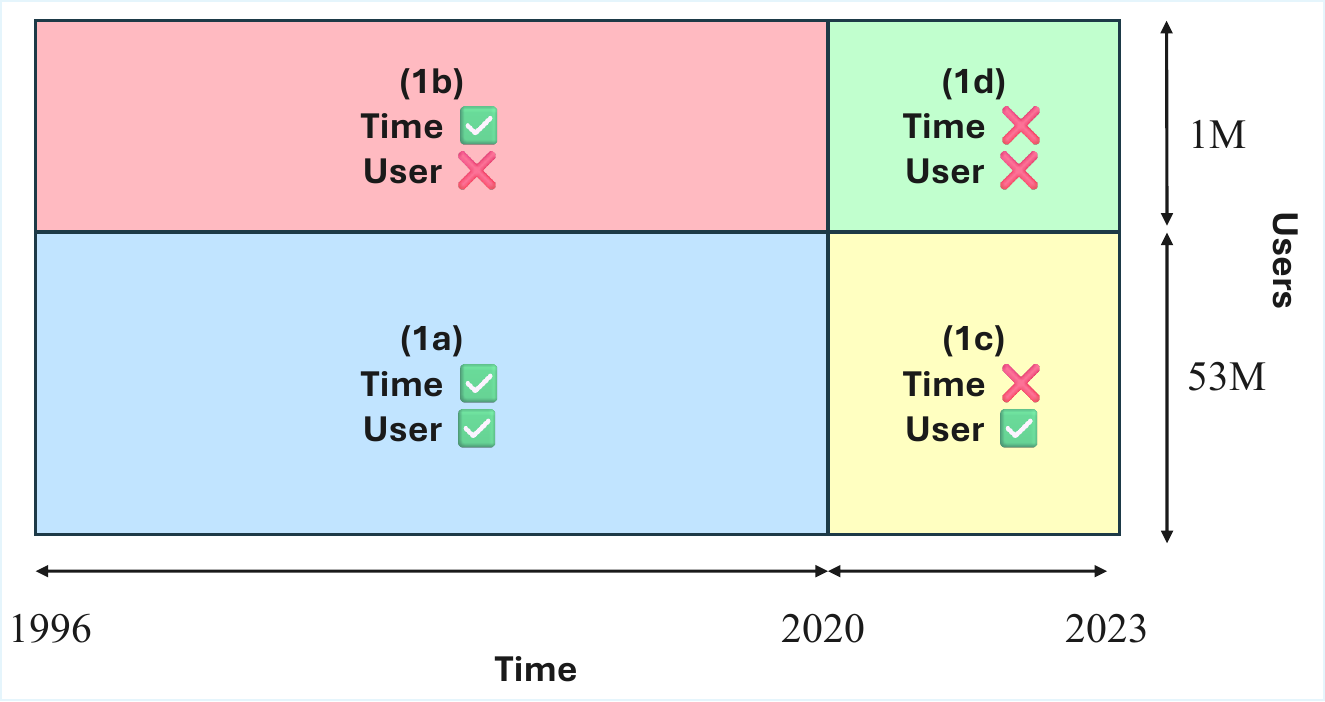}
  \caption{Proposed evaluation splits on the \texttt{HORIZON} benchmark for Task 1.}
  \label{fig:splits_expln}
\end{figure}

\noindent\textbf{Ratio-Based.} Here, the user sequence is split into training, validation, and test segments based on interaction timestamps, such that the resulting partition approximately adheres to a predefined ratio—typically $8{:}1{:}1$. While this method introduces some variability in sequence lengths and improves over the deterministic nature of Leave-One-Out, it fails to evaluate out-of-distribution generalization to unseen users. Moreover, it can result in temporal leakage, where interactions in the training set of one user may overlap in time with test interactions of another ~\citep{10.1145/3383313.3418479}.

To conclude, both paradigms fail to evaluate generalization under distribution shift, as validation and test sets largely resemble the training distribution. This undermines the assessment of model robustness and overlooks the temporal evolution of user preferences, which central to real-world user behavior modeling. To address these shortcomings, we propose a temporally grounded evaluation protocol that enforces a fixed time-based cutoff and includes held-out users to explicitly test for extrapolation and out-of-distribution generalization to unseen users.

\subsection{Proposed Task Formulations on \texttt{HORIZON}}
To address the aforementioned limitations, we introduce a multi-axis evaluation protocol on the \texttt{HORIZON} benchmark that disentangles generalization across time and user identity. This design enables a controlled analysis of each factor's impact. We instantiate this framework through three tasks, beginning with traditional next-item recommendation, followed by two LLM-based user modeling tasks motivated by their growing industrial relevance.

\paragraph{Task 1 — Next Item Recommendation.} 
In the traditional next-item recommendation setting, given a user history $\mathcal{H}_u = [i_1, i_2, \dots, i_t]$, the goal is to recommend the next likely item $i_{t+1}$. 
To move beyond static evaluation protocols, we adopt a \textit{global temporal cut-off} $\tau$, using interactions before $\tau$ for training and those after for evaluation. 
This preserves temporal order and induces realistic distribution shifts, following prior work that advocates temporally grounded evaluation in recommender systems \citep{10.1145/3383313.3418479}.  

Crucially, rather than treating temporal generalization in isolation, we explicitly factor evaluation along two orthogonal dimensions: \textit{temporal position} (before vs.\ after $\tau$) and \textit{user visibility} (users seen vs.\ unseen during training). 
This yields four complementary evaluation settings that systematically disentangle in-distribution performance, temporal extrapolation, user-level generalization, and their combination (illustrated in \Cref{fig:splits_expln}). 
Together, these settings provide a fine-grained view of model behavior under progressively harder and more realistic distribution shifts, while using a single, fixed training protocol\footnote{Analysis of Distribution Shifts in Appendix \ref{appendix:B}, \ref{appendix:C}}:

\begin{enumerate}
    \item[(1a)] \textbf{In-Distribution, Temporally Aligned Evaluation (Leave-One-Out).} 
    For users included in training, we hold out their final interaction \textit{before} the global cut-off $\tau$ for testing, using the preceding interactions for training. This mirrors the standard Leave-One-Out setup, but restricted to a temporally consistent subset. It evaluates short-context prediction under a distribution closely matched with training, and constitutes the \textbf{only training setup} we propose.

    \item[(1b)] \textbf{In-Distribution, Temporal Extrapolation Evaluation (All-Post-$\tau$).}  
    Using the model trained in 1a, we evaluate on the full sequence of interactions occurring \textit{after} the temporal cut-off $\tau$ for the same users. Ground-truth items are incrementally revealed during evaluation, enabling assessment of temporal generalization and user preference evolution.

    \item[(1c)] \textbf{OOD-User, Temporally Aligned Evaluation (Leave-One-Out).}  
    For held-out users unseen during training, we perform Leave-One-Out evaluation on interactions before the temporal cut-off $\tau$. This setting assesses generalization to new users under temporal conditions aligned with the training distribution.

    \item[(1d)] \textbf{OOD-User, Temporal Extrapolation Evaluation (All-Post-$\tau$).}  
    In this most challenging setting, the model predicts all interactions following the temporal cut-off $\tau$ for held-out users unseen during training. With short, temporally recent histories as input, this task evaluates generalization across both user identity and time.
\end{enumerate}





\paragraph{Task 2 - LLM-Based Next Item Recommendation:}  

\begin{figure*}[t]
    \centering
    \includegraphics[width=0.85\textwidth]{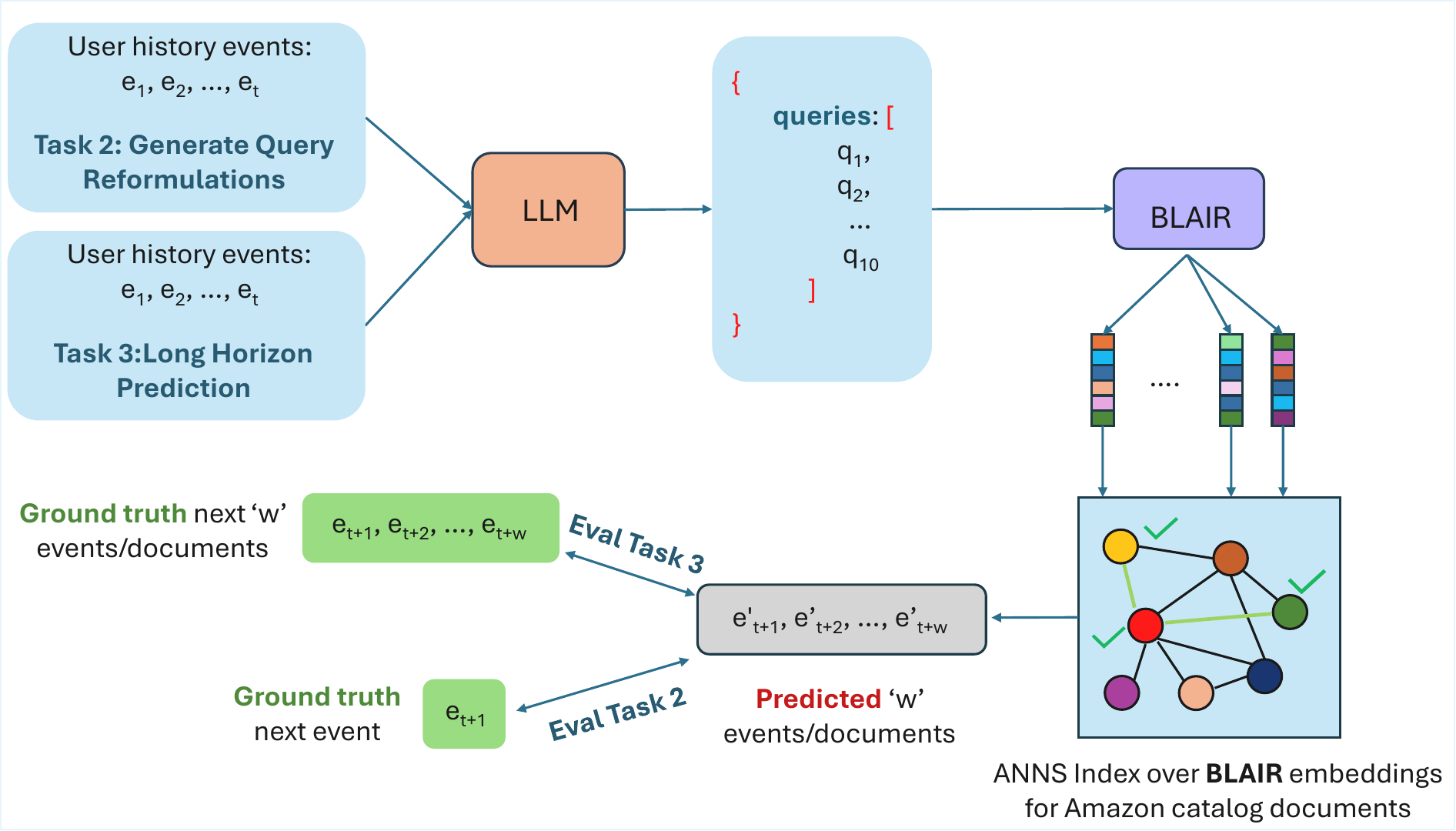}  
    \caption{Pipeline Detailing the LLM Generation, Retrieval and Evaluation Process Proposed for Tasks 2 and 3.}
    \label{fig:task2_3_formulation}
\end{figure*}


 
Large language models (LLMs), extensively pre-trained on web-scale corpora, are well-suited to model semantic patterns in text. As demonstrated in \Cref{fig:task2_3_formulation}, in this task, we treat the LLM as a user behavior encoder that reformulates a given user history into a diverse set of ten search queries $Q = \{q_1, \dots, q_{10}\}$, intended to capture various aspects of user intent and preference.\footnote{Prompts in Appendix \ref{appendix:prompt}} These queries are mapped, along with catalog items $i_j \in \mathcal{I}$, into a shared embedding space using a pre-trained item encoder.

An approximate nearest neighbor (ANN) index is constructed over catalog item embeddings $\{\mathbf{i}_j\}$, and top-$K$ candidates are retrieved for each query embedding $\mathbf{q}_k$. These are merged to form a final set of $K$ recommendations $\hat{I}$. We evaluate retrieval quality using Precision@K and Recall@K, based on cosine similarity with the ground truth item. 

Beyond ranking metrics, this task uniquely probes the LLM’s ability to generate high-quality, interpretable queries that reflect the underlying user behavior—serving as a semantic bridge between user history and candidate retrieval.

\paragraph{Task 3 - LLM-Based Long-Horizon User Modeling.}  
Traditional sequential recommendation typically focuses on predicting the immediate next item in a user’s interaction sequence. However, user modeling requires capturing longer-term behavior patterns that unfold over extended time windows \citep{zhou2024use, PinnerFormer}. Motivated by this, we propose the long-horizon modeling task on the \texttt{HORIZON} benchmark, leveraging the availability of longer and cross-domain user histories.

Given a user’s interaction history prior to a temporal cut-off, the LLM is tasked with generating natural language descriptions of the next $10$ items the user is likely to engage with. These descriptions represent a high-level summary of future behavior across a wider horizon. Using the same retrieval pipeline as Task 2, each generated description is embedded and used to retrieve matching items from the catalog. 

Evaluation is performed using standard retrieval metrics (e.g., Recall@K, Precision@K) by comparing the retrieved items with ground truth future interactions. This task assesses the LLM’s ability to model long-term user intent and will be an essential evaluation for assessing their ability to model user behaviors.

\section{Experimental Setup}
\label{sec:exp_setups}
\paragraph{Task 1 Setup:}
We adopt a temporal cut-off of $\tau = 2020$ to define the training window. From the full dataset of $\sim$54M users, we randomly sample 1M users who have any post-$\tau$ interactions as our out-of-distribution (OOD) user set, and treat the remaining 53M as the in-distribution (IND) pool. From this IND pool, 1M users are sampled to construct the test set for sub-task (1c). Due to computational constraints, we train all models on a 100K user subset of the IND set, and evaluate on 25K users each for sub-task (1d) (IND extrapolation) and sub-task (1c) (OOD prediction). \footnote{Detailed stats highlighting difference between the splits is added to the Appendix.} For all baselines, we use the \textsc{RecBole} framework~\citep{recbole,recbole2.0}, which offers standardized implementations and reproducible pipelines for recommendation models. The following popular item-ID-based baselines are included:

\textsc{GRU4Rec}~\citep{gru4rec} employs a recurrent architecture with gated recurrent units to capture sequential dependencies in user histories. \textsc{SASRec}~\citep{SASRec} adopts a transformer-based architecture with self-attention mechanisms to model user behavior sequences. \textsc{BERT4Rec}~\citep{BERT4Rec} utilizes a bidirectional transformer encoder trained with a Cloze-style objective to leverage full-sequence context. \textsc{CORE}~\citep{core} formulates session representations as weighted linear combinations of item embeddings, aligning both session and item vectors in a shared latent space.

While these methods are typically evaluated using either ratio-based or leave-one-out strategies, we retrain and evaluate them under the temporally grounded evaluation protocol described in the previous section. All models are trained with standardized hyperparameters and evaluated on our four evaluation settings using \textbf{MRR}, \textbf{Recall@K}, and \textbf{NDCG@K} for $K = \{10, 50, 100\}$.%

\begin{table*}[!htbp]
    \begin{minipage}[b]{0.48\linewidth}
    \centering
    \tiny
    \scriptsize
    \caption{In-Distribution User - Temporally Aligned Evaluation (N=NDCG, M=MRR, R=Recall)}
    \label{tab:in_Distribution_aligned}
    \setlength{\tabcolsep}{2pt}
    \begin{tabular}{l *{9}{c}}
        \toprule
        \textbf{Baseline} & \multicolumn{3}{c}{\textbf{N}} & \multicolumn{3}{c}{\textbf{M}} & \multicolumn{3}{c}{\textbf{R}} \\
        \cmidrule(lr){2-4} \cmidrule(lr){5-7} \cmidrule(lr){8-10}
        & 10 & 50 & 100 & 10 & 50 & 100 & 10 & 50 & 100 \\
        \midrule
        \textbf{CORE}     & 8.5  & 8.7  & 8.7  & 7.25 & 7.30 & 7.30 & 12.1 & 13.0 & 13.4 \\
        \textbf{SASRec}   & 25.2 & 27.4 & 27.9 & 22.5 & 22.9 & 23.0 & \textbf{34.1} & \textbf{43.8} & \textbf{46.6} \\
        \textbf{BERT4Rec} & \textbf{26.4} & \textbf{27.8} & \textbf{28.2} & \textbf{23.9} & \textbf{24.2} & \textbf{24.3} & 33.9 & 40.4 & 42.9 \\
        \textbf{GRU4Rec}  & 0.08 & 0.12 & 0.14 & 0.07 & 0.07 & 0.08 & 0.14 & 0.31 & 0.43 \\
        \bottomrule
    \end{tabular}
    \end{minipage}
    \tiny
    \hfill
    \begin{minipage}[b]{0.48\linewidth}
    \centering
    \scriptsize
    \caption{OOD User - Temporally Aligned Evaluation (N=NDCG, M=MRR, R=Recall)}
    \label{tab:ood_aligned}
    \setlength{\tabcolsep}{2pt}
    \begin{tabular}{l *{9}{c}}
        \toprule
        \textbf{Baseline} & \multicolumn{3}{c}{\textbf{N}} & \multicolumn{3}{c}{\textbf{M}} & \multicolumn{3}{c}{\textbf{R}} \\
        \cmidrule(lr){2-4} \cmidrule(lr){5-7} \cmidrule(lr){8-10}
        & 10 & 50 & 100 & 10 & 50 & 100 & 10 & 50 & 100 \\
        \midrule
        \textbf{CORE} & 5.9  & 6.8  & 7.2  & 4.19 & 4.39 & 4.43 & 11.1 & 15.4 & 17.9 \\
        \textbf{SASRec}   & \textbf{17.8} & \textbf{19.2} & \textbf{19.6} & \textbf{15.2} & \textbf{15.5} & \textbf{15.5} & \textbf{26.2} & \textbf{32.2} & 34.6 \\
        \textbf{BERT4Rec} & 11.8 & 14.4 & 15.2 & 9.96 & 10.50 & 10.58 & 17.8 & 29.5 & \textbf{34.7} \\
        \textbf{GRU4Rec}  & 0.01 & 0.01 & 0.02 & 0.004 & 0.004 & 0.005 & 0.01 & 0.03 & 0.08 \\
        \bottomrule
    \end{tabular}
    \end{minipage}
\end{table*}

\begin{table*}[!htbp]
\begin{minipage}[b]{0.48\linewidth}
    \tiny
    \centering
    \scriptsize
    \caption{In-Distribution User - Temporal Extrapolation Evaluation (N=NDCG, M=MRR, R=Recall)}
    \label{tab:in_distribution_extrapolation}
    \setlength{\tabcolsep}{2pt}
    \begin{tabular}{l *{9}{c}}
        \toprule
        \textbf{Baseline} & \multicolumn{3}{c}{\textbf{N}} & \multicolumn{3}{c}{\textbf{M}} & \multicolumn{3}{c}{\textbf{R}} \\
        \cmidrule(lr){2-4} \cmidrule(lr){5-7} \cmidrule(lr){8-10}
        & 10 & 50 & 100 & 10 & 50 & 100 & 10 & 50 & 100 \\
        \midrule
        \textbf{CORE} & 0.09 & 0.47 & 0.75 & 0.04 & 0.11 & 0.13 & 0.26 & 2.10 & 3.78 \\
        \textbf{SASRec} & \textbf{2.9}  & \textbf{3.6}  & 3.9  & \textbf{1.88} & \textbf{2.03} & \textbf{2.05} & \textbf{6.2}  & 9.4  & 11.0 \\
        \textbf{BERT4Rec} & 1.1  & 3.2  & \textbf{4.0}  & 0.56 & 0.99 & 1.10 & 2.8  & \textbf{12.8} & \textbf{17.8} \\
        \textbf{GRU4Rec}  & 0.01 & 0.01 & 0.02 & 0.004 & 0.005 & 0.01 & 0.01 & 0.03 & 0.08 \\
        \bottomrule
    \end{tabular}
\end{minipage}
\hfill
\begin{minipage}[b]{0.48\linewidth}
    \centering
    \tiny
    \scriptsize
    \caption{OOD User - Temporal Extrapolation Evaluation (N=NDCG, M=MRR, R=Recall)}
    \label{tab:ood_extrapolation}
    \setlength{\tabcolsep}{2pt}
    \begin{tabular}{l *{9}{c}}
        \toprule
        \textbf{Baseline} & \multicolumn{3}{c}{\textbf{N}} & \multicolumn{3}{c}{\textbf{M}} & \multicolumn{3}{c}{\textbf{R}} \\
        \cmidrule(lr){2-4} \cmidrule(lr){5-7} \cmidrule(lr){8-10}
        & 10 & 50 & 100 & 10 & 50 & 100 & 10 & 50 & 100 \\
        \midrule
        \textbf{CORE}     & 0.10 & 0.53 & 0.82 & 0.04 & 0.12 & 0.15 & 0.32 & 2.33 & 4.13 \\
        \textbf{SASRec}  & \textbf{3.1}  & \textbf{3.9}  & \textbf{4.1}  & \textbf{2.01} & \textbf{2.17} & \textbf{2.19} & \textbf{6.7}  & 9.9  & 11.6 \\
        \textbf{BERT4Rec} & 1.1  & 3.4  & 4.3  & 0.55 & 1.02 & 1.10 & 2.8  & \textbf{13.7} & \textbf{18.9} \\
        \textbf{GRU4Rec}  & 0.01 & 0.01 & 0.02 & 0.004 & 0.004 & 0.005 & 0.01 & 0.04 & 0.07 \\
        \bottomrule
    \end{tabular}
\end{minipage}
\end{table*}

\paragraph{Task 2 and 3 Setup:}
For Tasks 2 and 3, we use the held-out out-of-distribution (OOD) test set comprising 1M users as our evaluation benchmark. We primarily focus on evaluating the zero-shot capabilities of LLMs for modeling user behavior, as effective training paradigms for LLMs in recommendation settings remain an open research problem and present unique challenges in our context given the extremely long-tailed item distribution. Nevertheless, we include standard fine-tuning baselines (PEFT and full fine-tuning) for completeness. 

To ensure the integrity of comparisons, we avoid merging the in-distribution (IND) test set used in Task 1 with this OOD evaluation pool, as we use the same set for over fine-tuning baselines. We evaluate three recent and publicly available language models up to 9B parameter scale: \textsc{LLaMA-3.1-8B} \citep{grattafiori2024llama}, \textsc{Qwen3-8B} \citep{yang2025qwen3technicalreport}, and \textsc{Gemma2-9B} \citep{gemmateam2024gemma2improvingopen}. All models are queried in a zero-shot manner using a standardized prompt for each task. 

For encoding the items and queries, we use the pre-trained BLAIR item encoder \citep{hou2024bridging} as it is pre-trained on the Amazon-Reviews items and the FAISS library \citep{douze2024faiss} for creating the ANN-based vector databases to perform retrieval. 
As we do not perform ranking across queries, we compute standard retrieval metrics i.e. \textsc{Recall@K} and \textsc{Precision@K} for $K = 10, 50, 100$ to assess the effectiveness of the generated outputs in retrieving relevant items. \footnote{Detailed hyperparameter settings, prompts and implementation details are provided in the Appendix.}

\section{Results \& Discussion}
\begin{table*}[!htbp]
\begin{minipage}[b]{0.48\linewidth}
    \tiny
    \centering
    \setlength{\tabcolsep}{5pt}
    \caption{LLM-based Query Reformulation}
    \label{tab:query_reformulation}
    \begin{tabular}{l ccc ccc}
        \toprule
        \multirow{2}{*}{\textbf{Model}} & \multicolumn{3}{c}{\textbf{Recall}} & \multicolumn{3}{c}{\textbf{Precision}} \\
        \cmidrule(lr){2-4} \cmidrule(lr){5-7}
                                           & @10    & @50    & @100    & @10     & @50     & @100     \\
        \midrule        
        \textbf{LLAMA-3.1-8B} & 1.62 & 2.37 & 2.84 & 0.20 & 0.23 & 0.22 \\
        \textbf{Qwen3-8B} & \textbf{2.06} & \textbf{2.95} & \textbf{3.50} & \textbf{0.25} & \textbf{0.28} & \textbf{0.28} \\
        \textbf{Gemma2-9B} & 1.45 & 2.26 & 2.66 & 0.16 & 0.21 & 0.19 \\
        \bottomrule
    \end{tabular}
\end{minipage}%
\hfill
\begin{minipage}[b]{0.48\linewidth}
    \centering
    \tiny
    \setlength{\tabcolsep}{5pt}
    \caption{Long-Horizon User Modeling}
    \label{tab:long_horizon_modeling}
    \begin{tabular}{l ccc ccc}
        \toprule
        \multirow{2}{*}{\textbf{Model}} & \multicolumn{3}{c}{\textbf{Recall}} & \multicolumn{3}{c}{\textbf{Precision}} \\
        \cmidrule(lr){2-4} \cmidrule(lr){5-7}
                                           & @10    & @50    & @100    & @10     & @50     & @100     \\
        \midrule
        \textbf{LLAMA-3.1-8B} & 1.26 & 6.52 & 13.25 & 0.51 & 0.52 & 0.53 \\
        \textbf{Qwen3-8B} & \textbf{1.51} & \textbf{7.78} & \textbf{15.75} & \textbf{0.63} & \textbf{0.65} & \textbf{0.66} \\
        \textbf{Gemma2-9B} & 0.98 & 5.07 & 10.39 & 0.42 & 0.43 & 0.44 \\
        \bottomrule
    \end{tabular}
\end{minipage}
\end{table*}

\subsection{Benchmarking traditional ID-based baselines}
\Cref{tab:in_Distribution_aligned,tab:in_distribution_extrapolation,tab:ood_aligned,tab:ood_extrapolation} demonstrate the performance of traditional ID-based baselines across both In-Distribution as well as of OOD user settings across both temporal setups. 

\paragraph{Challenging Nature of the Task} Our task formulation is substantially more challenging than prior work such as ~\citep{hou2024bridging}. Unlike settings that train and evaluate on narrow category-wise domains and subsets (eg: Beauty), we train on the full distribution of user activity spanning diverse product categories, aiming to predict the next plausible item in a multi-domain environment. While non-attention-based models like \textsc{GRU4Rec}~\citep{gru4rec} have shown strong results in simpler setups~\citep{reprod_analysis}, we find that they struggle in our broader and more complex setting. In contrast, attention-based models such as \textsc{BERT4Rec, SASRec} perform markedly better, underscoring the importance of flexible context modeling.

\paragraph{Is traditional in-distribution leave-one-out evaluation sufficient?} 
Standard evaluation protocols in recommendation typically adopt an in-distribution leave-one-out setting, where a user's next interaction—already seen during training—is held out as the target. As shown in \Cref{tab:in_Distribution_aligned}, models achieve strong performance under this setup. However, in our more realistic out-of-distribution (OOD) evaluation, where test users are entirely unseen during training, we observe \textbf{a significant performance drop across all methods}. This highlights that in-distribution protocols may overestimate generalization and motivates the need for OOD-based evaluation to better assess true model robustness.

\paragraph{Is model performance stable on OOD users?}
Despite the degradation observed in OOD evaluation (\Cref{tab:ood_aligned}), attention-based models retain relatively high accuracy. This indicates that shared behavioral patterns across users; especially within the same temporal window can still be leveraged, consistent with collaborative filtering principles. While OOD evaluation poses a harder challenge, these results suggest that robust sequence modeling can partially bridge the generalization gap.

\paragraph{Impact of temporal distributional shifts} As shown in \Cref{tab:in_distribution_extrapolation,tab:ood_extrapolation}, temporal shifts result in significant performance degradation across all models, irrespective of whether the user was seen during training. Notably, models generalize better to unseen users from the same time period than to the same users across different time periods. We attribute this to the reliance on ID-based representations, which lack semantic grounding. Consequently, models struggle to adapt to new items, such as those from emerging brands, due to an absence of similarity encoding with previously seen items. This highlights the need for incorporating textual or semantic features to enhance robustness.

\subsection{Benchmarking LLM-based Query Reformulation for Recommendation} 
\Cref{tab:query_reformulation} summarizes the evaluation of three prominent LLMs i.e. LLAMA-3.1-8B, Qwen3-8B, and Gemma2-9B on their ability to reformulate user queries for item retrieval. The results indicate modest performance in both Recall and Precision, with improvements as the number of recommendations increases from 10 to 100. This trend suggests that LLMs capture some relevant items within larger candidate sets, reflecting a partial understanding of broader user intent. Nonetheless, absolute Recall values remain relatively low, indicating challenges in consistently retrieving a substantial portion of truly relevant items. Among the models, Qwen3-8B consistently outperforms LLAMA-3.1-8B and Gemma2-9B. 

To further evaluate query quality, we measured semantic similarity between reformulated queries and ground-truth items using BLAIR embeddings. The average cosine similarity scores are approximately 0.73 for Qwen3-8B, 0.72 for Gemma2-9B, and 0.71 for LLAMA-3.1-8B. These moderate similarities indicate that, while the queries capture reasonable semantic relatedness, there remains scope to generate better quality reformulations. 

We also evaluate reasoning models like Qwen3-235B in both reasoning and non-reasoning modes to understand if \textbf{Large Reasoning Models} can be an effective solution. We observe a \textbf{Recall@100 of 2.96} in Reasoning mode and \textbf{3.4} (Non-Reasoning mode), which is comparable in performance with the Qwen3-8B model, suggesting model scaling or reasoning may not be very effective currently. 

Lastly, we conducted \textbf{LLM fine-tuning experiments} using parameter-efficient (LoRA) \citep{hu2022lora} and full fine-tuning approaches with LLaMA-3.1-8B and Qwen3-8B models. Fine-tuned models are comparable to the zero-shot setting on next-item recommendation (best of 10 performance) (see \Cref{tab:llm_finetuning,tab:query_reformulation}), with the zero-shot approach being more scalable for growing catalogs. We elaborate on these results in \Cref{app:fine-tuning}.

\subsection{Benchmarking Results on Long-Horizon settings}
\Cref{tab:long_horizon_modeling} shows how LLMs perform on the challenging task of long-horizon user modeling, where models generate high-level summaries of future interests that are mapped to catalog items. While Recall improves with higher $k$, indicating some relevance capture over long sequences, Precision remains low, reflecting a high rate of irrelevant predictions. Among the models, Qwen3-8B consistently outperforms LLAMA-3.1-8B and Gemma2-9B. When comparing with the Query Reformulation task (\Cref{tab:query_reformulation}), it's important to note that long-horizon evaluation benefits from multiple ground-truth targets, unlike the single-reference setup in query reformulation. This may overstate long-horizon performance, even though it requires modeling deeper preference evolution. Furthermore, prior work often relaxes strict ordering in long-horizon evaluation ~\citep{PinnerFormer,zhou2024use}, further complicating direct metric comparisons. Thus, evaluations across these tasks should be interpreted with care, given their differing objectives.

\section{Conclusion}
\label{sec:Conclusion}
User behavior modeling is central to modern personalized systems, yet existing benchmarks and evaluation protocols fall short of capturing the generalization challenges faced in real-world deployments. In this work, we introduced \textsc{Horizon}, a benchmark designed to evaluate sequential recommendation models under realistic long-horizon, temporal, and user-level distribution shifts. 

\textsc{Horizon} defines five evaluation settings that systematically disentangle in-distribution performance, temporal extrapolation, and generalization to unseen users. Through extensive experiments across multiple model families, we demonstrate that model performance varies substantially across these settings, exposing limitations that are obscured by conventional evaluation practices. 

Our findings highlight the need for evaluation frameworks that move beyond short-horizon next-item accuracy and explicitly measure robustness under realistic generalization scenarios. By providing a large-scale, cross-domain benchmark and a principled evaluation protocol, \textsc{Horizon} aims to support the development of more robust user modeling methods for modern personalized platforms.

\section{Limitations}
\label{sec:limitations}
Despite the comprehensive nature of our benchmark, we acknowledge several limitations in our current work. Our benchmark is currently restricted to English-only data, representing an important opportunity for future extension to multilingual contexts. Additionally, the dataset is limited to textual modalities, though expanding to multimodal data would significantly enhance its applicability for multimodal user modeling tasks that integrate visual / auditory information. For our experiments using the RecBole framework, we were limited to using only a subset of the complete dataset for training and evaluation due to computational resource constraints. Lastly, the scope of our experimentation is restricted to the e-commerce setting. However, our evaluation framework is extensible to any arbitrary domain.
\section{Acknowledgements}
We sincerely thank Yashoteja Prabhu (Microsoft Research) and Medha Hira (Carnegie Mellon University) for their insightful discussions and generous review of this draft, which helped us organize our ideas more effectively and improve its overall readability. 

\bibliography{custom}

@article{hou2024bridging,
  title={Bridging language and items for retrieval and recommendation},
  author={Hou, Yupeng and Li, Jiacheng and He, Zhankui and Yan, An and Chen, Xiusi and McAuley, Julian},
  journal={arXiv preprint arXiv:2403.03952},
  year={2024}
}

@inproceedings{wu-etal-2020-mind,
    title = "{MIND}: A Large-scale Dataset for News Recommendation",
    author = "Wu, Fangzhao  and
      Qiao, Ying  and
      Chen, Jiun-Hung  and
      Wu, Chuhan  and
      Qi, Tao  and
      Lian, Jianxun  and
      Liu, Danyang  and
      Xie, Xing  and
      Gao, Jianfeng  and
      Wu, Winnie  and
      Zhou, Ming",
    editor = "Jurafsky, Dan  and
      Chai, Joyce  and
      Schluter, Natalie  and
      Tetreault, Joel",
    booktitle = "Proceedings of the 58th Annual Meeting of the Association for Computational Linguistics",
    month = jul,
    year = "2020",
    address = "Online",
    publisher = "Association for Computational Linguistics",
    url = "https://aclanthology.org/2020.acl-main.331/",
    doi = "10.18653/v1/2020.acl-main.331",
    pages = "3597--3606"
}

@inproceedings{10.1007/978-3-031-49601-1_4,
author = {Anand, Avinash and Goel, Arnav and Hira, Medha and Buldeo, Snehal and Kumar, Jatin and Verma, Astha and Gupta, Rushali and Shah, Rajiv Ratn},
title = {SciPhyRAG - Retrieval Augmentation to Improve LLMs on Physics Q\&A},
year = {2023},
isbn = {978-3-031-49600-4},
publisher = {Springer-Verlag},
address = {Berlin, Heidelberg},
url = {https://doi.org/10.1007/978-3-031-49601-1_4},
doi = {10.1007/978-3-031-49601-1_4},
booktitle = {Big Data and Artificial Intelligence: 11th International Conference, BDA 2023, Delhi, India, December 7–9, 2023, Proceedings},
pages = {50–63},
numpages = {14},
location = {Delhi, India}
}

@inproceedings{10.1145/3728483.3760194,
author = {Kapuriya, Janak and Shaikh, Anwar and Goel, Arnav and Hira, Medha and Singh, Apoorv and Saraf, Jay and Sanjana and Nauriyal, Vaibhav and Anand, Avinash and Wang, Zhengkui and Shah, Rajiv Ratn},
title = {Enhancing Scientific Visual Question Answering via Vision-Caption aware Supervised Fine-Tuning},
year = {2025},
isbn = {9798400718403},
publisher = {Association for Computing Machinery},
address = {New York, NY, USA},
url = {https://doi.org/10.1145/3728483.3760194},
doi = {10.1145/3728483.3760194},
booktitle = {Proceedings of the 2nd International Workshop on Large Vision - Language Model Learning and Applications},
pages = {13–30},
numpages = {18},
location = {Ireland},
series = {LAVA '25}
}

@misc{li2025attributingcultureconditionedgenerationspretraining,
      title={Attributing Culture-Conditioned Generations to Pretraining Corpora}, 
      author={Huihan Li and Arnav Goel and Keyu He and Xiang Ren},
      year={2025},
      eprint={2412.20760},
      archivePrefix={arXiv},
      primaryClass={cs.CL},
      url={https://arxiv.org/abs/2412.20760}, 
}

@misc{goel2023advancementsscientificcontrollabletext,
      title={Advancements in Scientific Controllable Text Generation Methods}, 
      author={Arnav Goel and Medha Hira and Avinash Anand and Siddhesh Bangar and Rajiv Ratn Shah},
      year={2023},
      eprint={2307.05538},
      archivePrefix={arXiv},
      primaryClass={cs.CL},
      url={https://arxiv.org/abs/2307.05538}, 
}

@inproceedings{10.1145/2020408.2020579,
author = {Cho, Eunjoon and Myers, Seth A. and Leskovec, Jure},
title = {Friendship and mobility: user movement in location-based social networks},
year = {2011},
isbn = {9781450308137},
publisher = {Association for Computing Machinery},
address = {New York, NY, USA},
url = {https://doi.org/10.1145/2020408.2020579},
doi = {10.1145/2020408.2020579},
booktitle = {Proceedings of the 17th ACM SIGKDD International Conference on Knowledge Discovery and Data Mining},
pages = {1082–1090},
numpages = {9},
location = {San Diego, California, USA},
series = {KDD '11}
}

@inproceedings{10.1145/3383313.3418479,
author = {Meng, Zaiqiao and McCreadie, Richard and Macdonald, Craig and Ounis, Iadh},
title = {Exploring Data Splitting Strategies for the Evaluation of Recommendation Models},
year = {2020},
isbn = {9781450375832},
publisher = {Association for Computing Machinery},
address = {New York, NY, USA},
url = {https://doi.org/10.1145/3383313.3418479},
doi = {10.1145/3383313.3418479},
booktitle = {Proceedings of the 14th ACM Conference on Recommender Systems},
pages = {681–686},
numpages = {6},
location = {Virtual Event, Brazil},
series = {RecSys '20}
}

@article{grattafiori2024llama,
  title={The llama 3 herd of models},
  author={Grattafiori, Aaron and Dubey, Abhimanyu and Jauhri, Abhinav and Pandey, Abhinav and Kadian, Abhishek and Al-Dahle, Ahmad and Letman, Aiesha and Mathur, Akhil and Schelten, Alan and Vaughan, Alex and others},
  journal={arXiv preprint arXiv:2407.21783},
  year={2024}
}

@misc{gemmateam2024gemma2improvingopen,
      title={Gemma 2: Improving Open Language Models at a Practical Size}, 
      author={Gemma Team and Morgane Riviere and Shreya Pathak and Pier Giuseppe Sessa and Cassidy Hardin and Surya Bhupatiraju and Léonard Hussenot and Thomas Mesnard and Bobak Shahriari and Alexandre Ramé and Johan Ferret and Peter Liu and Pouya Tafti and Abe Friesen and Michelle Casbon and Sabela Ramos and Ravin Kumar and Charline Le Lan and Sammy Jerome and Anton Tsitsulin and Nino Vieillard and Piotr Stanczyk and Sertan Girgin and Nikola Momchev and Matt Hoffman and Shantanu Thakoor and Jean-Bastien Grill and Behnam Neyshabur and Olivier Bachem and Alanna Walton and Aliaksei Severyn and Alicia Parrish and Aliya Ahmad and Allen Hutchison and Alvin Abdagic and Amanda Carl and Amy Shen and Andy Brock and Andy Coenen and Anthony Laforge and Antonia Paterson and Ben Bastian and Bilal Piot and Bo Wu and Brandon Royal and Charlie Chen and Chintu Kumar and Chris Perry and Chris Welty and Christopher A. Choquette-Choo and Danila Sinopalnikov and David Weinberger and Dimple Vijaykumar and Dominika Rogozińska and Dustin Herbison and Elisa Bandy and Emma Wang and Eric Noland and Erica Moreira and Evan Senter and Evgenii Eltyshev and Francesco Visin and Gabriel Rasskin and Gary Wei and Glenn Cameron and Gus Martins and Hadi Hashemi and Hanna Klimczak-Plucińska and Harleen Batra and Harsh Dhand and Ivan Nardini and Jacinda Mein and Jack Zhou and James Svensson and Jeff Stanway and Jetha Chan and Jin Peng Zhou and Joana Carrasqueira and Joana Iljazi and Jocelyn Becker and Joe Fernandez and Joost van Amersfoort and Josh Gordon and Josh Lipschultz and Josh Newlan and Ju-yeong Ji and Kareem Mohamed and Kartikeya Badola and Kat Black and Katie Millican and Keelin McDonell and Kelvin Nguyen and Kiranbir Sodhia and Kish Greene and Lars Lowe Sjoesund and Lauren Usui and Laurent Sifre and Lena Heuermann and Leticia Lago and Lilly McNealus and Livio Baldini Soares and Logan Kilpatrick and Lucas Dixon and Luciano Martins and Machel Reid and Manvinder Singh and Mark Iverson and Martin Görner and Mat Velloso and Mateo Wirth and Matt Davidow and Matt Miller and Matthew Rahtz and Matthew Watson and Meg Risdal and Mehran Kazemi and Michael Moynihan and Ming Zhang and Minsuk Kahng and Minwoo Park and Mofi Rahman and Mohit Khatwani and Natalie Dao and Nenshad Bardoliwalla and Nesh Devanathan and Neta Dumai and Nilay Chauhan and Oscar Wahltinez and Pankil Botarda and Parker Barnes and Paul Barham and Paul Michel and Pengchong Jin and Petko Georgiev and Phil Culliton and Pradeep Kuppala and Ramona Comanescu and Ramona Merhej and Reena Jana and Reza Ardeshir Rokni and Rishabh Agarwal and Ryan Mullins and Samaneh Saadat and Sara Mc Carthy and Sarah Cogan and Sarah Perrin and Sébastien M. R. Arnold and Sebastian Krause and Shengyang Dai and Shruti Garg and Shruti Sheth and Sue Ronstrom and Susan Chan and Timothy Jordan and Ting Yu and Tom Eccles and Tom Hennigan and Tomas Kocisky and Tulsee Doshi and Vihan Jain and Vikas Yadav and Vilobh Meshram and Vishal Dharmadhikari and Warren Barkley and Wei Wei and Wenming Ye and Woohyun Han and Woosuk Kwon and Xiang Xu and Zhe Shen and Zhitao Gong and Zichuan Wei and Victor Cotruta and Phoebe Kirk and Anand Rao and Minh Giang and Ludovic Peran and Tris Warkentin and Eli Collins and Joelle Barral and Zoubin Ghahramani and Raia Hadsell and D. Sculley and Jeanine Banks and Anca Dragan and Slav Petrov and Oriol Vinyals and Jeff Dean and Demis Hassabis and Koray Kavukcuoglu and Clement Farabet and Elena Buchatskaya and Sebastian Borgeaud and Noah Fiedel and Armand Joulin and Kathleen Kenealy and Robert Dadashi and Alek Andreev},
      year={2024},
      eprint={2408.00118},
      archivePrefix={arXiv},
      primaryClass={cs.CL},
      url={https://arxiv.org/abs/2408.00118}, 
}

@misc{yang2025qwen3technicalreport,
      title={Qwen3 Technical Report}, 
      author={An Yang and Anfeng Li and Baosong Yang and Beichen Zhang and Binyuan Hui and Bo Zheng and Bowen Yu and Chang Gao and Chengen Huang and Chenxu Lv and Chujie Zheng and Dayiheng Liu and Fan Zhou and Fei Huang and Feng Hu and Hao Ge and Haoran Wei and Huan Lin and Jialong Tang and Jian Yang and Jianhong Tu and Jianwei Zhang and Jianxin Yang and Jiaxi Yang and Jing Zhou and Jingren Zhou and Junyang Lin and Kai Dang and Keqin Bao and Kexin Yang and Le Yu and Lianghao Deng and Mei Li and Mingfeng Xue and Mingze Li and Pei Zhang and Peng Wang and Qin Zhu and Rui Men and Ruize Gao and Shixuan Liu and Shuang Luo and Tianhao Li and Tianyi Tang and Wenbiao Yin and Xingzhang Ren and Xinyu Wang and Xinyu Zhang and Xuancheng Ren and Yang Fan and Yang Su and Yichang Zhang and Yinger Zhang and Yu Wan and Yuqiong Liu and Zekun Wang and Zeyu Cui and Zhenru Zhang and Zhipeng Zhou and Zihan Qiu},
      year={2025},
      eprint={2505.09388},
      archivePrefix={arXiv},
      primaryClass={cs.CL},
      url={https://arxiv.org/abs/2505.09388}, 
}

@inproceedings{steamdataset,
author = {Sobkowicz, Antoni and Stokowiec, Wojciech},
year = {2016},
month = {05},
pages = {},
title = {Steam Review Dataset - new, large scale sentiment dataset}
}

@article{douze2024faiss,
  title={The faiss library},
  author={Douze, Matthijs and Guzhva, Alexandr and Deng, Chengqi and Johnson, Jeff and Szilvasy, Gergely and Mazar{\'e}, Pierre-Emmanuel and Lomeli, Maria and Hosseini, Lucas and J{\'e}gou, Herv{\'e}},
  journal={arXiv preprint arXiv:2401.08281},
  year={2024}
}

@inproceedings{amazon-m2,
author = {Jin, Wei and Mao, Haitao and Li, Zheng and Jiang, Haoming and Luo, Chen and Wen, Hongzhi and Han, Haoyu and Lu, Hanqing and Wang, Zhengyang and Li, Ruirui and Li, Zhen and Cheng, Monica and Goutam, Rahul and Zhang, Haiyang and Subbian, Karthik and Wang, Suhang and Sun, Yizhou and Tang, Jiliang and Yin, Bing and Tang, Xianfeng},
title = {Amazon-M2: a multilingual multi-locale shopping session dataset for recommendation and text generation},
year = {2023},
publisher = {Curran Associates Inc.},
address = {Red Hook, NY, USA},
booktitle = {Proceedings of the 37th International Conference on Neural Information Processing Systems},
articleno = {351},
numpages = {21},
location = {New Orleans, LA, USA},
series = {NIPS '23}
}

@article{movielens,
author = {Harper, F. Maxwell and Konstan, Joseph A.},
title = {The MovieLens Datasets: History and Context},
year = {2015},
issue_date = {January 2016},
publisher = {Association for Computing Machinery},
address = {New York, NY, USA},
volume = {5},
number = {4},
issn = {2160-6455},
url = {https://doi.org/10.1145/2827872},
doi = {10.1145/2827872},
journal = {ACM Trans. Interact. Intell. Syst.},
month = dec,
articleno = {19},
numpages = {19}
}

@inproceedings{sun2023take,
  title={Take a fresh look at recommender systems from an evaluation standpoint},
  author={Sun, Aixin},
  booktitle={Proceedings of the 46th International ACM SIGIR Conference on Research and Development in Information Retrieval},
  pages={2629--2638},
  year={2023}
}

@article{ji2023critical,
  title={A critical study on data leakage in recommender system offline evaluation},
  author={Ji, Yitong and Sun, Aixin and Zhang, Jie and Li, Chenliang},
  journal={ACM Transactions on Information Systems},
  volume={41},
  number={3},
  pages={1--27},
  year={2023},
  publisher={ACM New York, NY}
}

@article{mcelfresh2022generalizability,
  title={On the generalizability and predictability of recommender systems},
  author={McElfresh, Duncan and Khandagale, Sujay and Valverde, Jonathan and Dickerson, John and White, Colin},
  journal={Advances in Neural Information Processing Systems},
  volume={35},
  pages={4416--4432},
  year={2022}
}

@inproceedings{wu2020mind,
  title={Mind: A large-scale dataset for news recommendation},
  author={Wu, Fangzhao and Qiao, Ying and Chen, Jiun-Hung and Wu, Chuhan and Qi, Tao and Lian, Jianxun and Liu, Danyang and Xie, Xing and Gao, Jianfeng and Wu, Winnie and others},
  booktitle={Proceedings of the 58th annual meeting of the association for computational linguistics},
  pages={3597--3606},
  year={2020}
}

@inproceedings{core,
author = {Hou, Yupeng and Hu, Binbin and Zhang, Zhiqiang and Zhao, Wayne Xin},
title = {CORE: Simple and Effective Session-based Recommendation within Consistent Representation Space},
year = {2022},
isbn = {9781450387323},
publisher = {Association for Computing Machinery},
address = {New York, NY, USA},
url = {https://doi.org/10.1145/3477495.3531955},
doi = {10.1145/3477495.3531955},
booktitle = {Proceedings of the 45th International ACM SIGIR Conference on Research and Development in Information Retrieval},
pages = {1796–1801},
numpages = {6},
location = {Madrid, Spain},
series = {SIGIR '22}
}

@misc{gru4rec,
      title={Session-based Recommendations with Recurrent Neural Networks}, 
      author={Balázs Hidasi and Alexandros Karatzoglou and Linas Baltrunas and Domonkos Tikk},
      year={2016},
      eprint={1511.06939},
      archivePrefix={arXiv},
      primaryClass={cs.LG},
      url={https://arxiv.org/abs/1511.06939}, 
}

@INPROCEEDINGS{SASRec,
  author={Kang, Wang-Cheng and McAuley, Julian},
  booktitle={2018 IEEE International Conference on Data Mining (ICDM)}, 
  title={Self-Attentive Sequential Recommendation}, 
  year={2018},
  volume={},
  number={},
  pages={197-206},
  doi={10.1109/ICDM.2018.00035}}

@article{jin2023amazon,
  title={Amazon-m2: A multilingual multi-locale shopping session dataset for recommendation and text generation},
  author={Jin, Wei and Mao, Haitao and Li, Zheng and Jiang, Haoming and Luo, Chen and Wen, Hongzhi and Han, Haoyu and Lu, Hanqing and Wang, Zhengyang and Li, Ruirui and others},
  journal={Advances in Neural Information Processing Systems},
  volume={36},
  pages={8006--8026},
  year={2023}
}

@inproceedings{BERT4Rec,
author = {Sun, Fei and Liu, Jun and Wu, Jian and Pei, Changhua and Lin, Xiao and Ou, Wenwu and Jiang, Peng},
title = {BERT4Rec: Sequential Recommendation with Bidirectional Encoder Representations from Transformer},
year = {2019},
isbn = {9781450369763},
publisher = {Association for Computing Machinery},
address = {New York, NY, USA},
url = {https://doi.org/10.1145/3357384.3357895},
doi = {10.1145/3357384.3357895},
booktitle = {Proceedings of the 28th ACM International Conference on Information and Knowledge Management},
pages = {1441–1450},
numpages = {10},
location = {Beijing, China},
series = {CIKM '19}
}

@inproceedings{PinnerFormer,
author = {Pancha, Nikil and Zhai, Andrew and Leskovec, Jure and Rosenberg, Charles},
title = {PinnerFormer: Sequence Modeling for User Representation at Pinterest},
year = {2022},
isbn = {9781450393850},
publisher = {Association for Computing Machinery},
address = {New York, NY, USA},
url = {https://doi.org/10.1145/3534678.3539156},
doi = {10.1145/3534678.3539156},
booktitle = {Proceedings of the 28th ACM SIGKDD Conference on Knowledge Discovery and Data Mining},
pages = {3702–3712},
numpages = {11},
location = {Washington DC, USA},
series = {KDD '22}
}

@ARTICLE{reprod_analysis,
  author={Betello, Filippo and Purificato, Antonio and Siciliano, Federico and Trappolini, Giovanni and Bacciu, Andrea and Tonellotto, Nicola and Silvestri, Fabrizio},
  journal={IEEE Access}, 
  title={A Reproducible Analysis of Sequential Recommender Systems}, 
  year={2025},
  volume={13},
  number={},
  pages={5762-5772},
  doi={10.1109/ACCESS.2024.3522049}}

@article{harper2015movielens,
  title={The movielens datasets: History and context},
  author={Harper, F Maxwell and Konstan, Joseph A},
  journal={Acm transactions on interactive intelligent systems (tiis)},
  volume={5},
  number={4},
  pages={1--19},
  year={2015},
  publisher={Acm New York, NY, USA}
}

@inproceedings{ni2019justifying,
  title={Justifying recommendations using distantly-labeled reviews and fine-grained aspects},
  author={Ni, Jianmo and Li, Jiacheng and McAuley, Julian},
  booktitle={Proceedings of the 2019 conference on empirical methods in natural language processing and the 9th international joint conference on natural language processing (EMNLP-IJCNLP)},
  pages={188--197},
  year={2019}
}

@article{zhou2024use,
  title={USE: Dynamic User Modeling with Stateful Sequence Models},
  author={Zhou, Zhihan and Fang, Qixiang and Neves, Leonardo and Barbieri, Francesco and Liu, Yozen and Liu, Han and Bos, Maarten W and Dotsch, Ron},
  journal={arXiv preprint arXiv:2403.13344},
  year={2024}
}

@article{treves2025viki,
  title={VIKI: Systematic Cross-Platform Profile Inference of Online Users},
  author={Treves, Ben and De Cristofaro, Emiliano and Dong, Yue and Faloutsos, Michalis},
  journal={arXiv preprint arXiv:2503.14772},
  year={2025}
}

@inproceedings{hou2022towards,
  title={Towards universal sequence representation learning for recommender systems},
  author={Hou, Yupeng and Mu, Shanlei and Zhao, Wayne Xin and Li, Yaliang and Ding, Bolin and Wen, Ji-Rong},
  booktitle={Proceedings of the 28th ACM SIGKDD Conference on Knowledge Discovery and Data Mining},
  pages={585--593},
  year={2022}
}

@inproceedings{lin2019cross,
  title={Cross: Cross-platform recommendation for social e-commerce},
  author={Lin, Tzu-Heng and Gao, Chen and Li, Yong},
  booktitle={Proceedings of the 42nd International ACM SIGIR conference on research and development in information retrieval},
  pages={515--524},
  year={2019}
}

@inproceedings{recbole,
    author={Wayne Xin Zhao and Shanlei Mu and Yupeng Hou and Zihan Lin and Yushuo Chen and Xingyu Pan and Kaiyuan Li and Yujie Lu and Hui Wang and Changxin Tian and Yingqian Min and Zhichao Feng and Xinyan Fan and Xu Chen and Pengfei Wang and Wendi Ji and Yaliang Li and Xiaoling Wang and Ji{-}Rong Wen},
    title={RecBole: Towards a Unified, Comprehensive and Efficient Framework for Recommendation Algorithms},
    booktitle={{CIKM}},
    pages={4653--4664},
    publisher={{ACM}},
    year={2021}
 }

@article{recbole2.0,
    author={Wayne Xin Zhao and Yupeng Hou and Xingyu Pan and Chen Yang and Zeyu Zhang and Zihan Lin and Jingsen Zhang and Shuqing Bian and Jiakai Tang and Wenqi Sun and Yushuo Chen and Lanling Xu and Gaowei Zhang and Zhen Tian and Changxin Tian and Shanlei Mu and Xinyan Fan and Xu Chen and Ji{-}Rong Wen},
    title={RecBole 2.0: Towards a More Up-to-Date Recommendation Library},
    journal={arXiv preprint arXiv:2206.07351},
    year={2022}
 }

@inproceedings{rendle2020neural,
  title={Neural collaborative filtering vs. matrix factorization revisited},
  author={Rendle, Steffen and Krichene, Walid and Zhang, Li and Anderson, John},
  booktitle={Proceedings of the 14th ACM Conference on Recommender Systems},
  pages={240--248},
  year={2020}
}

@article{yue2023llamarec,
  title={Llamarec: Two-stage recommendation using large language models for ranking},
  author={Yue, Zhenrui and Rabhi, Sara and Moreira, Gabriel de Souza Pereira and Wang, Dong and Oldridge, Even},
  journal={arXiv preprint arXiv:2311.02089},
  year={2023}
}

@inproceedings{wang2024can,
  title={Can small language models be good reasoners for sequential recommendation?},
  author={Wang, Yuling and Tian, Changxin and Hu, Binbin and Yu, Yanhua and Liu, Ziqi and Zhang, Zhiqiang and Zhou, Jun and Pang, Liang and Wang, Xiao},
  booktitle={Proceedings of the ACM Web Conference 2024},
  pages={3876--3887},
  year={2024}
}

@inproceedings{klenitskiy2024we,
author = {Klenitskiy, Anton and Volodkevich, Anna and Pembek, Anton and Vasilev, Alexey},
title = {Does It Look Sequential? An Analysis of Datasets for Evaluation of Sequential Recommendations},
year = {2024},
isbn = {9798400705052},
publisher = {Association for Computing Machinery},
address = {New York, NY, USA},
url = {https://doi.org/10.1145/3640457.3688195},
doi = {10.1145/3640457.3688195},
booktitle = {Proceedings of the 18th ACM Conference on Recommender Systems},
pages = {1067–1072},
numpages = {6},
location = {Bari, Italy},
series = {RecSys '24}
}

@inproceedings{
hu2022lora,
title={Lo{RA}: Low-Rank Adaptation of Large Language Models},
author={Edward J Hu and Yelong Shen and Phillip Wallis and Zeyuan Allen-Zhu and Yuanzhi Li and Shean Wang and Lu Wang and Weizhu Chen},
booktitle={International Conference on Learning Representations},
year={2022},
url={https://openreview.net/forum?id=nZeVKeeFYf9}
}

\appendix
\section{\horizon\ Statistics and Plots} \label{appendix:A}

\begin{figure}[htbp]
    \centering
    \includegraphics[width=0.5\textwidth]{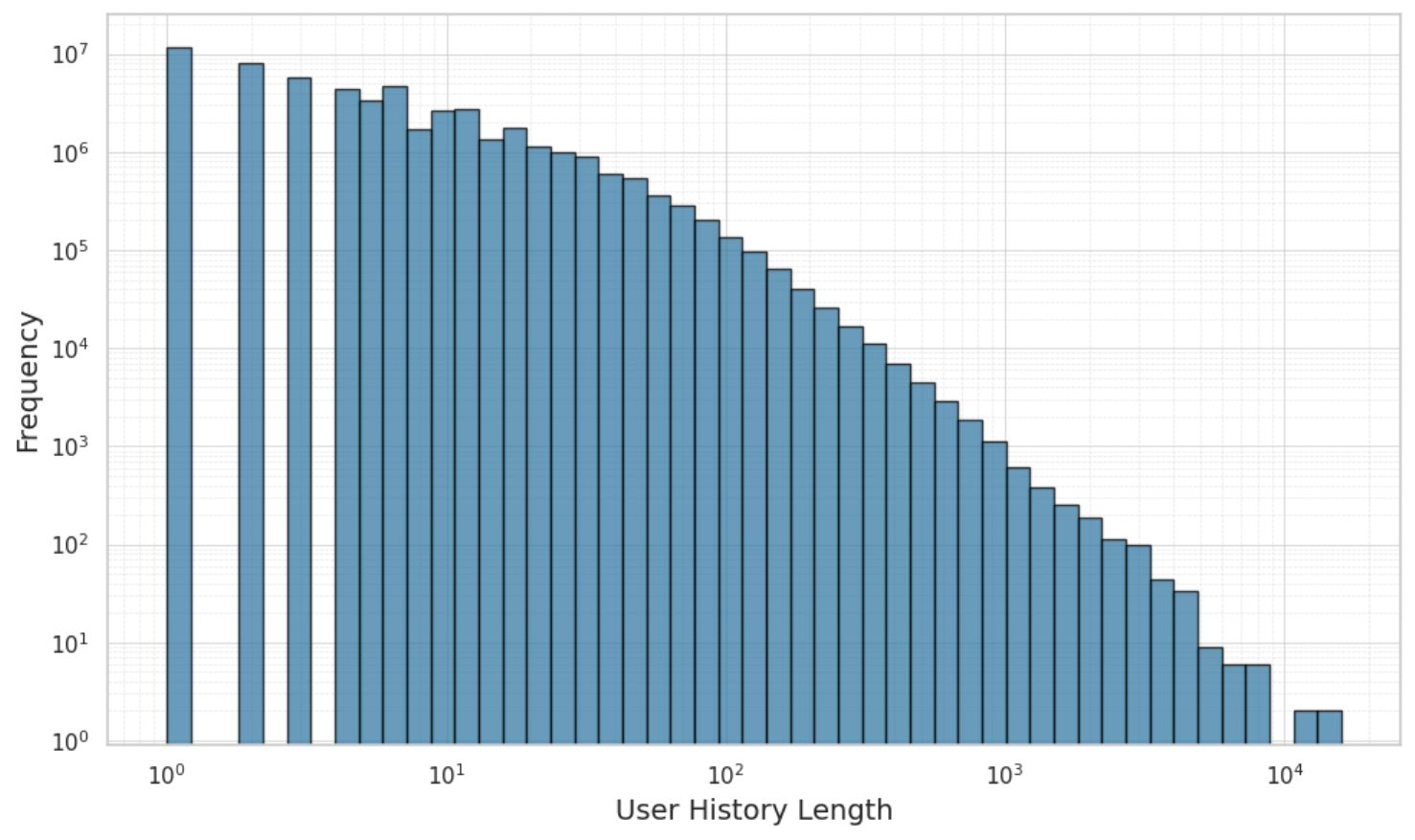}  
    \caption{Histogram Depicting the Frequency Distribution of User History Lengths in \horizon. The presence of ultra-long user histories highlights the need for architectures capable of modeling long-range sequential dependencies.}
    \label{fig:length_distribution}
\end{figure}

The \horizon\ benchmark is curated by reformulating the widely-used Amazon Reviews 2023 dataset \citep{hou2024bridging}, merging all 33 categories into unified user histories to enable robust long-term, cross-domain user modeling. This section provides an in-depth statistical analysis of the dataset through visualizations and derived insights.

\noindent\textbf{Scale and Diversity:} The benchmark comprises approximately \textbf{53.5M users and 34.5M unique items}, amounting to nearly 486M interaction records. This scale is significantly larger than prior public benchmarks and captures highly diverse behavioral patterns. With the unified formulation, user histories naturally span multiple product categories—introducing heterogeneous context that is both semantically diverse and temporally rich. This setting reflects real-world personalization challenges more faithfully than isolated category-based modeling.

\noindent\textbf{User History Lengths:} \Cref{fig:length_distribution} illustrates a long-tailed distribution of user history lengths in \horizon. While a large portion of users exhibit short interaction sequences, there exists a substantial number with extremely long histories—extending beyond 1,000 timestamps for tens of thousands of users. This highlights the need for models capable of handling long-range dependencies and memory-efficient representations. Traditional sequence models struggle in this regime due to vanishing gradients and computational bottlenecks, motivating the exploration of transformer-based or memory-augmented architectures for this benchmark.

\begin{figure}[htbp]
    \centering
    \includegraphics[width=0.46\textwidth]{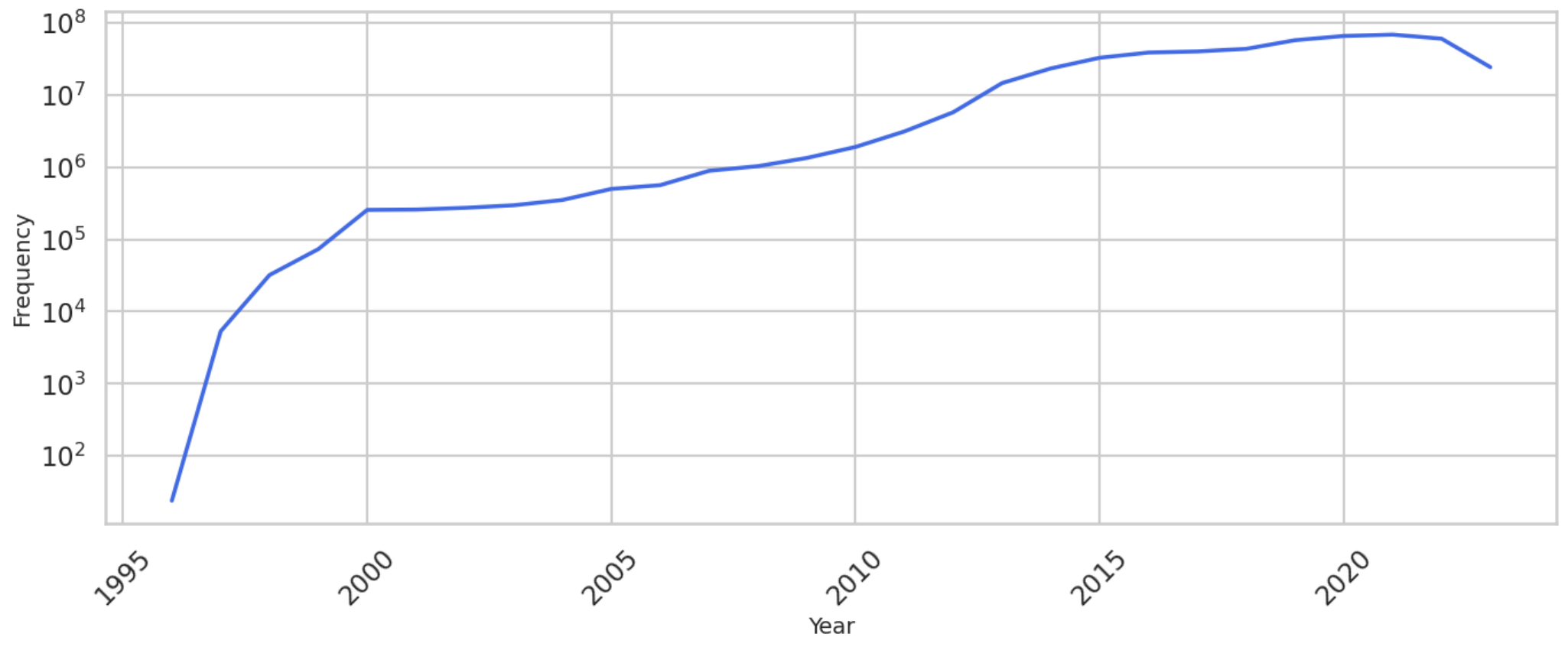}  
    \caption{Line Plot Depicting the Temporal Distribution of User Histories in \horizon. The balanced volume before and after 2020 makes it suitable for temporal extrapolation tasks.}
    \label{fig:temp_distribution}
\end{figure}

\noindent\textbf{Temporal Structure and Generalization:} The temporal distribution of interactions (\Cref{fig:temp_distribution}) reveals a sharp rise in user activity post-2010, peaking around 2020. Crucially, nearly half the interactions occur after the 2020 temporal cut-off used in our evaluation framework. Specifically, the average number of timestamps before 2020 is 4.99, while it remains comparable after 2020 at 4.09. This temporal balance ensures that both training and test splits are adequately rich, setting up a robust testbed for extrapolative evaluation and temporal generalization. As models are evaluated on unseen user interactions post-2020, they are challenged to infer future behavior patterns from past, potentially outdated, preferences—mirroring real-world drift in user intent.

\begin{figure*}[htbp]
    \centering
    \includegraphics[width=0.8\textwidth]{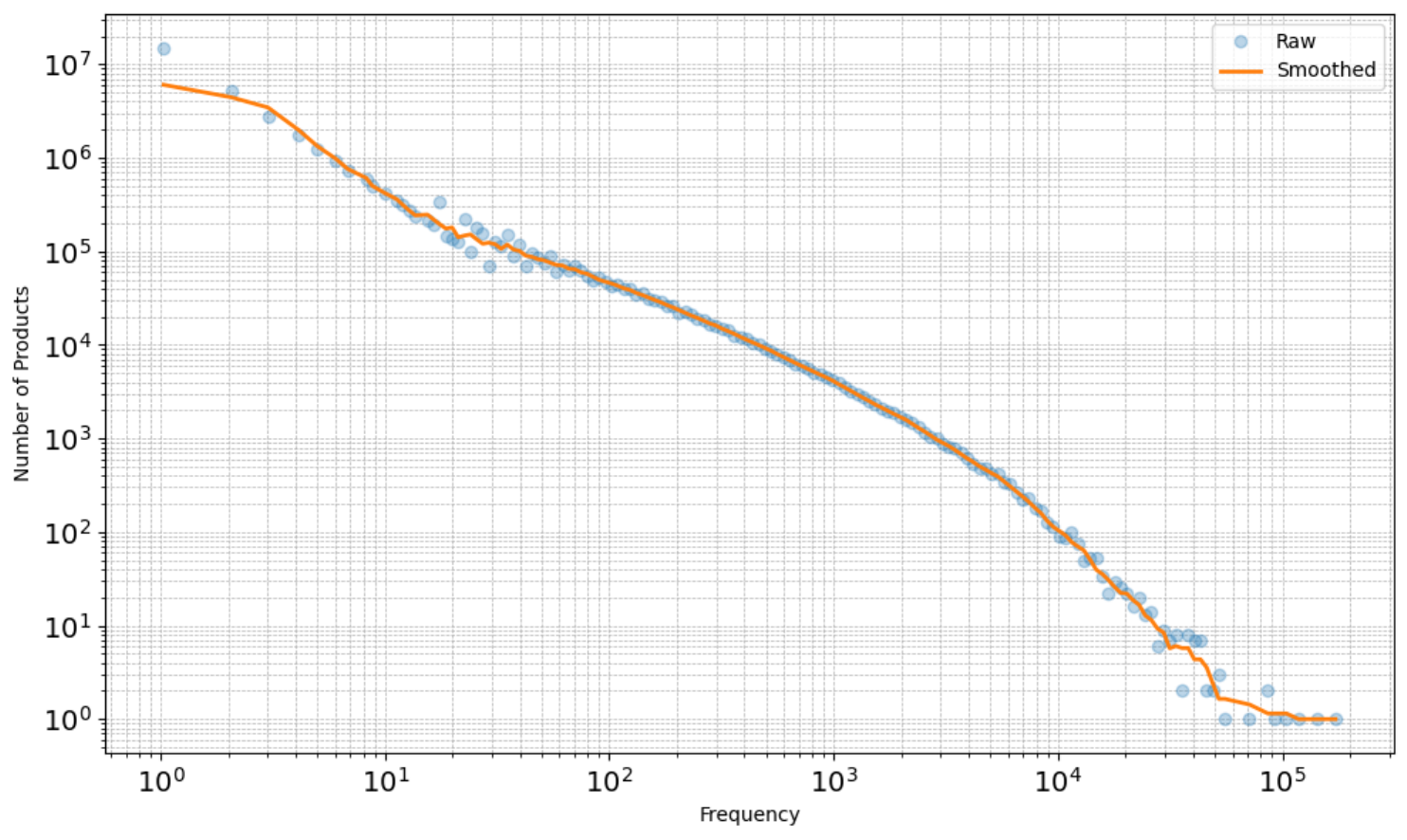}  
    \caption{Frequency Distribution of Products in the \horizon\ Benchmark. The power-law structure reflects extreme item sparsity, with most items having very few interactions.}
    \label{fig:product_id_dist}
\end{figure*}

\noindent\textbf{Product Distribution:} \Cref{fig:product_id_dist} plots the frequency distribution of product IDs, which exhibits a pronounced long-tail trend. A small fraction of items dominate interactions, while the majority are sparsely interacted with. This reflects typical e-commerce dynamics but poses unique challenges for recommender systems: most prior models are biased toward frequent items. The high item cardinality (34M) and sparse tail necessitate models that generalize well to rarely seen or previously unseen products. Incorporating textual features or content-based augmentations could be beneficial in this context, especially under cold-start settings.

\noindent\textbf{Benchmark Design Implications:} The three key observations from these plots underscore the difficulty of the \horizon\ benchmark:
\begin{enumerate}
    \item \textbf{Long Histories:} Users with thousands of interaction points require models that capture dependencies over extended horizons and adapt across evolving interests.
    \item \textbf{Temporal Drift:} A significant portion of test data lies beyond the training horizon (post-2020), enforcing extrapolation beyond the training distribution and testing robustness to temporal shifts.
    \item \textbf{Item Sparsity:} The skewed product frequency implies that many test items are low-frequency or unseen, further intensifying the generalization challenge.
\end{enumerate}

Taken together, \horizon\ enables a comprehensive stress test of user behavior models across multiple axes—scale, history length, temporal generalization, and sparsity. Its unified multi-category formulation fosters the development of general-purpose, temporally robust, and cross-domain recommendation architectures.

\section{Task 1 Splits and Out-of-Distribution Analysis} \label{appendix:B}

In our proposed Task 1 setup, the user population is explicitly partitioned into two cohorts to rigorously test generalization: \textit{in-distribution (IND)} users observed during training, and \textit{out-of-distribution (OOD)} users who are entirely held out. The fixed temporal cutoff at $\tau = 2020$ allows us to decouple user generalization from temporal extrapolation. Below, we elaborate on the statistical and structural distinctions between these cohorts, which underline the difficulty of the proposed evaluation.

\noindent\textbf{Temporal Shift and Behavioral Drift:} As visualized in \Cref{fig:temp_distribution}, a significant volume of user activity in the dataset occurs post-2020. By construction, OOD users are sampled from this post-2020 pool, whereas IND users have interactions both before and after the temporal boundary. This creates a natural distributional shift: the OOD cohort is inherently more recent and behaviorally different, reflecting newer products, evolving user preferences, and potentially different session structures. Hence, even under temporally aligned evaluation (Subtask 1c), the OOD test set exhibits non-trivial variance from the training distribution.

\noindent\textbf{Semantic Divergence via Topic Modeling.}
To investigate the semantic distinctiveness between in-distribution (IND) and out-of-distribution (OOD) user groups, we apply Latent Dirichlet Allocation (LDA) to model topics from user review histories, treating each user as a document composed of concatenated item descriptions and metadata. The resulting topic distributions uncover meaningful divergence in user interests. 

Both groups engage with broad product themes (e.g., books, electronics, fashion), yet OOD users demonstrate stronger focus on niche and emergent categories. For example, OOD-specific topics include terms like \textit{“telescope,” “kite,” “bjj,” “freemason,”} and musical instruments such as \textit{“guitar,” “ukulele,” “pedal”}, suggesting a shift toward specialized or subcultural interests. 

In contrast, IND topics reflect more mainstream and diversified engagement, including wellness supplements (e.g., \textit{“nootropic,” “creatine,” “arginine”}) and general home goods. To quantify these patterns, we compute entropy and dominance over user topic distributions. OOD users show significantly lower entropy (mean = 1.18 vs. 1.28) and higher topic dominance (mean = 0.51 vs. 0.48), indicating more focused topical preferences. A t-SNE projection of user topic vectors reveals clear separation between IND and OOD clusters. 

Additionally, the average KL divergence from IND to OOD topic distributions exceeds 0.8, reinforcing the semantic shift. These findings suggest that OOD generalization reflects not just temporal drift but substantive thematic changes in user behavior and product engagement.

\section{Cross-domain Statistics and Distribution Shifts} \label{appendix:C}

\begin{figure*}[htbp]
    \centering
    \includegraphics[width=0.8\textwidth]{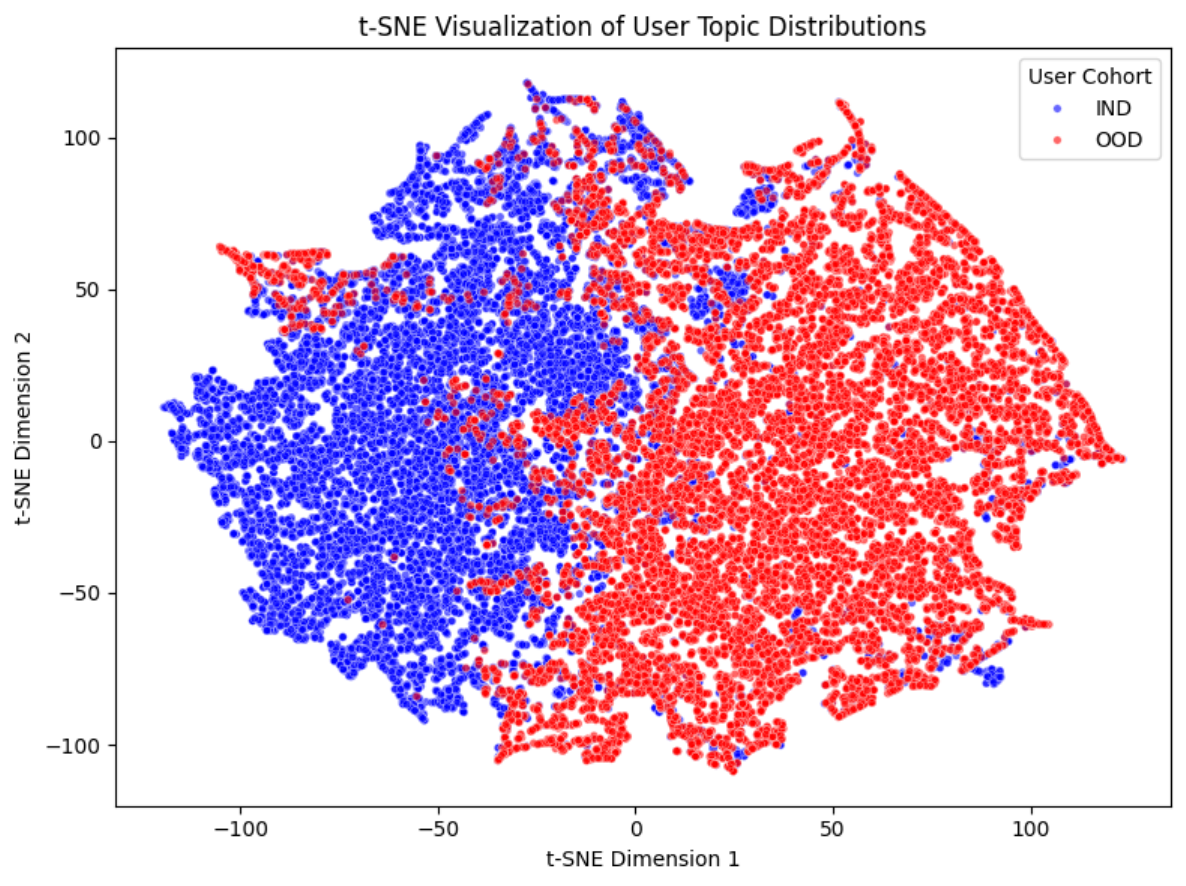}  
    \caption{t-SNE depicting the distinct user topic distributions in the in-distribution and OOD users.}
    \label{fig:t-sne_dist}
\end{figure*}

\paragraph{Unified cross-domain user histories.} 
Unlike category-segmented or short-timespan variants of Amazon Reviews, \textsc{Horizon} constructs fully unified user interaction sequences spanning all 35M items across diverse product domains. This design removes artificial boundaries between categories, allowing models to observe and learn from naturally interleaved behaviors (Section~3.1, Table~1). In practice, real-world users frequently transition between domains—e.g., electronics, books, and household items—within a single behavioral trajectory. By preserving this structure, \textsc{Horizon} enables the study of cross-domain preference transfer, long-range dependencies, and heterogeneous context modeling at unprecedented scale. Such evaluation is fundamentally infeasible in existing benchmarks such as MovieLens, MIND, or session-based datasets, which are either restricted to a single domain or limited to short temporal windows that truncate long-term behavioral signals.

\paragraph{Cross-domain user behavior under distribution shift.} 
A defining characteristic of \textsc{Horizon} is the prevalence of multi-domain user activity. On average, users interact with approximately six distinct domains, and over 90\% of users exhibit cross-domain behavior. This diversity becomes even more pronounced in the out-of-distribution (OOD) split, which is explicitly constructed to reflect more complex and heterogeneous user patterns. In particular, only 1\% of users in the OOD split remain single-domain, compared to 11\% in the training set. This shift indicates that models must generalize from relatively simpler, less diverse training trajectories to significantly richer and more entangled behavioral patterns at test time. As a result, \textsc{Horizon} naturally induces a distribution shift not only in item space but also in user behavior complexity, requiring models to capture transferable representations across domains rather than relying on narrow, domain-specific signals.

\paragraph{Item novelty, sparsity, and temporal dynamics.} 
The benchmark further introduces realistic challenges through a combination of item sparsity, temporal spread, and catalog evolution. OOD users tend to have sparse yet long-horizon interaction histories, with 63\% exhibiting between 6 and 20 interactions distributed across multiple years. This creates a setting where models must reason over temporally distant signals while operating under limited per-user data. Additionally, item exposure in the OOD split reflects significant novelty: 33\% of test interactions involve rare or tail items, nearly doubling the 19\% observed during training. Only 60\% of candidate items overlap with the training catalog, indicating substantial item turnover and the presence of previously unseen or underrepresented items at inference time. Together, these factors simulate realistic recommendation environments where user preferences evolve, item distributions shift, and long-tail discovery becomes critical—conditions that are largely absent in prior benchmarks such as MIND, Amazon-M2, and category-partitioned Amazon Reviews.

\section{Hyperparameters and Implementation Details}

\begin{table*}[htbp]
\centering
\caption{Model-specific hyperparameter configurations}
\label{tab:recbole_model_hyperparameters}
\begin{tabular}{|l|c|c|c|c|}
\hline
\textbf{Parameter} & \textbf{BERT4Rec} & \textbf{GRU4Rec} & \textbf{SASRec} & \textbf{CORE} \\
\hline
Hidden/Embedding size & 256 & 256 & 256 & 256 \\
Number of layers & 3 & 3 & 3 & 3 \\
Attention heads & 4 & - & 4 & 4 \\
Dropout probability & 0.15 & 0.15 & 0.15 & 0.15 \\
Batch size & 8192 & 8192 & 4096 & 4096 \\
Loss function & BPR & BPR & CE & CE \\
Mask ratio & 0.2 & - & - & - \\
\hline
\end{tabular}
\end{table*}

\subsection{RecBole Experiments - Task 1}

All models in Task 1 were trained using the RecBole framework \citep{recbole, recbole2.0} with a consistent configuration to ensure a fair comparison. The common training hyperparameters were selected based on prior literature and empirical tuning on a held-out validation set. These include a small learning rate of $2 \times 10^{-5}$ to stabilize optimization over long sequences, a maximum of 10 epochs for training, and early stopping with a patience of 10 epochs to prevent overfitting. To ensure reproducibility across all experimental runs, we fixed the random seed to 2025.

\paragraph{Training Hyperparameters.}
All models were trained with the following consistent configuration
\begin{itemize}
    \item \textbf{Learning rate}: $2 \times 10^{-5}$
    \item \textbf{Maximum epochs}: 10
    \item \textbf{Early stopping patience}: 10
    \item \textbf{Random seed}: 2025
    \item \textbf{Maximum sequence length}: 100
    \item \textbf{Validation metric}: MRR@10
    \item \textbf{Evaluation cutoffs}: $k \in \{10, 20, 50, 100\}$
    \item \textbf{Test negative samples}: 100
\end{itemize}

To support uniform evaluation across models, we truncated all user interaction sequences to a maximum of 100 items and used mean reciprocal rank at cutoff 10 (MRR@10) as the primary validation metric. During testing, we sampled 100 negative items for each user-item query to simulate realistic top-$k$ recommendation settings and report metrics at various cutoffs ($k$).

\paragraph{Model-Specific Hyperparameters}

Each model was configured using a 256-dimensional embedding and three layers to capture higher-order dependencies. Attention-based models (BERT4Rec, SASRec, and CORE) used 4 attention heads to balance modeling capacity and memory cost. A dropout rate of 0.15 was applied to all models for regularization. Batch sizes were tuned based on GPU memory availability and empirical training stability: 8192 for BERT4Rec and GRU4Rec, and 4096 for SASRec and CORE due to their higher per-batch memory footprint. These are further detailed in Table \ref{tab:recbole_model_hyperparameters}.

\paragraph{Architecture Details:} Given below are the architectural details about the RecBole baselines which we have employed in our study on the \horizon\ benchmark:
\begin{itemize}
    \item \textbf{BERT4Rec}: It leverages bidirectional Transformers to model sequence-wide context and predicts masked items using a masked language modeling (MLM) objective, with a mask ratio set to 0.2.
    \item \textbf{GRU4Rec}: GRU4Rec uses gated recurrent units (GRUs) to model sequential dependencies.
    \item \textbf{SASRec}: SASRec is built on unidirectional self-attention layers, enabling it to capture short- and long-term dependencies without recurrence.
    \item \textbf{CORE}: CORE integrates self-attention with collaborative filtering signals, enhancing personalization through a hybrid architecture
\end{itemize}

\paragraph{Loss Function Configuration.}
Given below are the possible loss function configurations available in RecBole for training sequential recommendation models:
\begin{itemize}
    \item \textbf{BPR models} (BERT4Rec, GRU4Rec, SASRec): Bayesian Personalized Ranking with negative sampling during training
    \item \textbf{CE models} (CORE): Cross-entropy loss without negative sampling during training
\end{itemize}

Models trained with BPR loss (BERT4Rec, GRU4Rec, SASRec) rely on dynamic negative sampling and optimize the ranking of positive over negative interactions. In contrast, CORE optimizes a classification objective using cross-entropy loss computed over the full softmax distribution.

\begin{table*}[h]
\centering
\caption{Hyperparameters used for different models.}
\begin{tabular}{lccc}
\toprule
\textbf{Hyperparameter} & \textsc{LLaMA-3.1-8B} & \textsc{Qwen3-8B} & \textsc{Gemma2-9B} \\
\midrule
Batch Size              & 512             & 512             & 256             \\
Temperature           & 0.7            & 0.7            & 0.7            \\
Top-P            & 0.8            & 0.8           & 0.8            \\
Top-K            & 20             & 20             & 20             \\
Max-Tokens (Task 2)             & 220             & 220            & 220             \\
Max-Tokens (Task 3)              & 350               & 350               & 350              \\
\bottomrule
\end{tabular}
\label{tab:hyperparams_llms}
\end{table*}

\paragraph{Execution Details.}
All experiments were conducted using a high-performance compute cluster equipped with 4 NVIDIA A100 GPUs (80GB VRAM each). We employed PyTorch’s automatic mixed precision (AMP) to accelerate training and reduce memory usage. Training time per epoch varied with architectural complexity: GRU4Rec, being lightweight, completed one epoch in approximately 0.75 hours, while BERT4Rec, with its attention-heavy encoder and MLM objective, required around 1.25 hours per epoch. Multi-GPU training was implemented using the NCCL backend for synchronized distributed training. All hyperparameters and implementation choices were fixed across all splits to ensure experimental consistency and comparability.

\subsection{Task 2 and 3 Experiments}

\paragraph{LLM Inference Setup.}
We adopt a consistent inference pipeline for both Task 2 (LLM-based Next Product Recommendation via Query Reformulation) and Task 3 (LLM-based Long-Horizon User Modeling), as described in Section 5 and illustrated in Figure 2. All models are prompted in a zero-shot setting, without any fine-tuning or retrieval augmentation, to evaluate their general-purpose reasoning capabilities over long user histories and contexts.

We utilize three state-of-the-art, instruction-tuned open-source LLMs: \textsc{LLaMA-3.1-8B} \citep{grattafiori2024llama}, \textsc{Qwen3-8B} \citep{yang2025qwen3technicalreport}, and \textsc{Gemma2-9B} \citep{gemmateam2024gemma2improvingopen}. These models were selected for their strong instruction-following capabilities and competitive performance on public benchmarks.

Table  \Cref{tab:hyperparams_llms} summarizes the decoding hyperparameters used. The temperature was fixed at 0.7 across all models to balance determinism and diversity in outputs. We set Top-P and Top-K sampling parameters based on model-specific best practices to control generation randomness. The maximum token limits were adjusted per task: 220 tokens for Task 2 (shorter search queries), and 350 tokens for Task 3 (longer next-item descriptions). Batch sizes were selected based on each model’s memory footprint and throughput on A100 GPUs, with the larger \textsc{Gemma2-9B} model using a smaller batch size.

\begin{table*}[!htbp]
\small
\centering
\caption{Comparison of Fine-tuned LLMs for Next-Item Prediction}
\label{tab:llm_finetuning}
\begin{tabular}{l|ccc|ccc}
\toprule
\multirow{2}{*}{\textbf{Setting}} & \multicolumn{3}{c|}{\textbf{Recall@K (\%)}} & \multicolumn{3}{c}{\textbf{Precision@K (\%)}} \\
\cmidrule(lr){2-4} \cmidrule(lr){5-7}
& \textbf{FFT (LLaMA3)} & \textbf{LoRA (LLaMA3)} & \textbf{LoRA (Qwen3)} & \textbf{FFT (LLaMA3)} & \textbf{LoRA (LLaMA3)} & \textbf{LoRA (Qwen3)} \\
\midrule
\multicolumn{7}{c}{\textit{In-Domain Temporal Extrapolation (Task 1c)}} \\
\midrule
K=10 & 1.45 & \textbf{1.65} & 1.38 & 0.98 & \textbf{1.29} & 0.90 \\
K=50 & 1.67 & \textbf{1.82} & 1.60 & 0.97 & \textbf{1.28} & 0.90 \\
K=100 & 2.02 & \textbf{2.09} & 1.93 & 0.97 & \textbf{1.28} & 0.89 \\
\midrule
\multicolumn{7}{c}{\textit{Out-of-Domain Temporal Extrapolation (Task 1d)}} \\
\midrule
K=10 & \textbf{1.24} & 0.71 & 1.18 & \textbf{0.82} & 0.42 & 0.77 \\
K=50 & \textbf{1.41} & 0.84 & 1.37 & \textbf{0.81} & 0.42 & 0.77 \\
K=100 & \textbf{1.71} & 1.07 & 1.67 & \textbf{0.80} & 0.42 & 0.76 \\
\bottomrule
\end{tabular}
\end{table*}
\paragraph{Execution Details.}
All inference was run using the vLLM engine on a compute cluster with 4× NVIDIA A100 40GB GPUs. The full test set consists of 1 million users, with each user processed independently in batched decoding mode. End-to-end inference across all models required approximately 5 days due to the volume of input prompts and the autoregressive nature of generation.

To support reproducibility and accessibility, we will release all evaluation code, prompt templates, and precomputed predictions on smaller held-out test splits post-acceptance. These subsets will enable low-resource experimentation on the same evaluation protocol without requiring access to large-scale GPU compute.

\paragraph{Generating Query and Item Embeddings using BLAIR.}
To encode the item catalog and predicted queries, we leverage the BLAIR item encoder \citep{hou2024bridging}, a RoBERTa-based model pretrained on Amazon review titles. We use the \texttt{hyp1231/blair-roberta-base} checkpoint via the HuggingFace Transformers library \footnote{\url{https://huggingface.co/hyp1231/blair-roberta-base}}, and tokenize each product title with a maximum sequence length of 512 tokens. Embeddings are obtained by extracting the \texttt{[CLS]} token representation from the final hidden layer, followed by $\ell_2$ normalization to facilitate cosine similarity-based retrieval. To scale embedding computation across a large number of titles, we utilize the \texttt{Accelerate} library with mixed-precision inference (fp16) and distributed processing across multiple GPUs, achieving efficient batch-wise encoding with a batch size of 4096. We shard the workload across processes and later merge the outputs to form a single embedding matrix for the catalog and prediction sets.

\paragraph{Retrieval and Indexing using FAISS.}
For approximate nearest neighbor (ANN) search, we employ the FAISS library \citep{douze2024faiss}, which implements the Hierarchical Navigable Small World (HNSW) graph-based indexing algorithm. We build a HNSW index on the catalog embeddings using cosine similarity as the distance metric. The key hyperparameters used during index construction include: \texttt{M=64}, which controls the number of bi-directional links created for each new node and influences index accuracy and memory usage; and \texttt{efConstruction=256}, which sets the dynamic list size for the graph during construction and affects indexing time and final recall quality. At query time, we use \texttt{efSearch=256} to control the breadth of the search and balance between latency and retrieval performance. These values were selected based on a grid search over the validation set to optimize top-$k$ recall, where $k=10$, while ensuring sub-millisecond retrieval latency per query on a modern GPU setup.

This setup enables scalable, low-latency nearest neighbor search over millions of product titles, while maintaining semantic alignment between predicted queries and candidate items.

\section{LLM-Finetuning Results} \label{app:fine-tuning}

\paragraph{What is the overall performance of fine-tuned LLMs?} 
\Cref{tab:llm_finetuning} reports the performance of fine-tuned LLMs for next-item prediction under temporal extrapolation, evaluated on both in-domain (Task 1c) and out-of-domain users (Task 1d). Across all configurations, performance remains uniformly low. In the in-domain setting, Recall@10 ranges between 1.38\% and 1.65\%, increasing only marginally to a maximum of 2.09\% at Recall@100 for LoRA fine-tuned LLaMA-3.1-8B. Precision follows a similarly flat trend, peaking at 1.29\%. Notably, increasing the cutoff from $K=10$ to $K=100$ yields only modest gains (e.g., 1.65\% $\rightarrow$ 2.09\% for LoRA LLaMA-3.1-8B), indicating that relevant items are not substantially concentrated even within larger candidate sets. 

Across training strategies, parameter-efficient fine-tuning (LoRA) provides slight but consistent improvements over full fine-tuning (FFT) for LLaMA-3.1-8B in the in-domain setting, though the absolute gains remain small (e.g., +0.07 at Recall@100). Qwen3-8B exhibits comparable but slightly weaker performance overall. These results suggest that neither scaling parameter updates nor switching architectures meaningfully improves next-item prediction under this formulation.

\paragraph{How does performance change under distribution shift?} 
Performance degrades consistently in the out-of-domain setting (Task 1d), where models must generalize to unseen users with more diverse and complex interaction patterns. Recall@10 drops from 1.65\% (LoRA LLaMA-3.1-8B, in-domain) to 0.71\% under OOD conditions, and Recall@100 decreases from 2.09\% to 1.07\%. A similar degradation is observed across all models, though the extent varies: FFT LLaMA-3.1-8B retains relatively stronger performance (1.71\% at Recall@100), while LoRA-based adaptations exhibit larger drops. Precision mirrors this behavior, remaining below 0.82\% across all configurations and dropping sharply for LoRA LLaMA-3.1-8B (0.42\% across all $K$). 

Another notable pattern is the reduced sensitivity to $K$ in the OOD setting. For example, LoRA LLaMA-3.1-8B increases only from 0.71\% to 1.07\% when moving from Recall@10 to Recall@100, compared to a larger relative gain in-domain. This suggests that under distribution shift, relevant items are not only harder to rank highly but also less likely to appear even in extended candidate lists. Overall, these results highlight the difficulty of extrapolating across both temporal shifts and user distributions when models are trained to generate a single next item from a large candidate space.

\paragraph{Do fine-tuned LLMs outperform simple baselines?} 
Despite the additional training, fine-tuned models achieve performance comparable to our zero-shot retrieval-based baseline (\Cref{tab:query_reformulation}). The retrieval approach generates ten semantically diverse queries per user without any task-specific optimization, yet matches the effectiveness of fine-tuned LLMs across both in-domain and OOD settings. Given the extremely low absolute recall values observed after fine-tuning, this parity is particularly striking. It suggests that instruction fine-tuning, when framed as single-item generation over extremely large and long-tailed item spaces, provides limited practical benefit. The consistency of this observation across architectures (LLaMA-3.1-8B and Qwen3-8B) and adaptation strategies further indicates that the limitation is not model-specific.

\paragraph{Why do fine-tuned LLMs struggle in this setting?} 
A key limitation lies in the training formulation. Unlike discriminative sequential recommenders, which leverage contrastive supervision and large-scale negative sampling to structure the item space, instruction-tuned LLMs operate under a purely generative objective. This provides no explicit signal to distinguish among millions of plausible candidates or to calibrate ranked outputs. The weak scaling of recall with increasing $K$, together with uniformly low precision, indicates that models fail to meaningfully organize the candidate space, often producing semantically plausible but misaligned predictions. Consequently, fine-tuning does not fully exploit the semantic and representational strengths of LLMs in long-tailed recommendation settings.

\paragraph{What directions could address these limitations?} 
These findings motivate alternative training paradigms that better align language modeling with recommendation objectives. Promising directions include contrastive or hybrid objectives with explicit negative sampling, as well as approaches that align model vocabularies with item identifiers for more direct reasoning over discrete item spaces. Additionally, multi-candidate generation or retrieval-augmented decoding may help bridge generative modeling with ranking-based evaluation, improving effectiveness under distribution shift.

\section{Prompts} \label{appendix:prompt}
\subsection{Task 2: Query-Based Next-Item Recommendation.}
As outlined in Section 4.2, this task evaluates an LLM’s ability to generate personalized search queries from a user's Amazon product history. The prompt asks the model to produce 10 queries balancing relevance with serendipity. These queries act as soft proxies for next-item prediction and reveal how well the model generalizes user intent. The setup is zero-shot, requiring the model to function as a semantic encoder-decoder without fine-tuning.

\begin{promptbox}
PROMPT FOR TASK 2 - LLM-Based Next Item Recommendation:

You are an expert at turning a user's Amazon product history into personalized search queries.

History: I1 <SEP> I2 <SEP> ..... <In>
This was the user's Amazon product history.

Your task is to generate a set of 10 personalized search queries that reflect the user's interests and preferences. 
Try to balance diversity and serendipity with relevancy to the user history. These queries will be used to recommend the next product to the user.

Out of these 10 queries:
4 queries should be directly related to the user's history;
3 queries should be tangentially related;
3 queries should be completely unrelated but interesting.

Process:
1. Think of a guideline explaining what intents or aspects you observed in the user history which helped you formulate these queries. You don't need to specify which is which.
2. Then, generate exactly 10 search queries balancing core interests with a bit of serendipity.

## Output Format
Provide the response only as a JSON object with one field: (do not generate anything else)

{
  "queries": [
    "query1",
    "query2",
    "...",
    "query10"
  ]
}
\end{promptbox}

\subsection{Task 3}
As described in Section 4.2, the following is the prompt for Task 3: Long-Horizon User Modeling using Large Language Models (LLMs). This task is designed to evaluate a model's ability to understand and extrapolate from a user's product history over time. The prompt guides the LLM to generate forward-looking, autoregressive item descriptions based on prior purchases, simulating realistic recommendation scenarios. Specifically, it instructs the model to infer underlying user preferences and behavioral patterns, and to generate coherent, temporally ordered predictions that balance relevance and serendipity. The prompt is framed in a zero-shot setting, encouraging the LLM to reason sequentially without access to explicit training examples.

\begin{promptbox}
PROMPT for Task 3 - LLM-Based Long-Horizon User Modeling:

You are an expert at predicting the next products a user may want based on their Amazon product history.

History: I1 <SEP> I2 <SEP> ..... <In>
This was the user's Amazon product history with exact product titles (NOT descriptions).

Your task is to generate descriptions for the next 10 items the user is most likely to be interested in. Provide concise, one - sentence descriptions that capture the essence of each potential item. These will guide recommendation generation.

Try to model the sequences in the user history and provide a mix of relevant and serendipitous items trying to capture the user's interests, intents and changes in behavior. Use the first item description to guide your next timestep's item description generation in an autoregressive manner.

Process:
1. Think of a guideline explaining the patterns or preferences you observed in the user history that informed your item descriptions.
2. Provide exactly 10 next-item descriptions balancing relevance and serendipity generated one after the other in temporal order.

## Output Format
Provide the response only as a JSON object with one field: (do not generate anything else)

{
  "item_descriptions_timewise": [
    "item_description_time_step1",
    "item_description_time_step2",
    "...",
    "item_description_time_step10"
  ]
}
\end{promptbox}

\end{document}